\RequirePackage{fix-cm}
\documentclass[smallextended]{svjour3}       
\smartqed  
\usepackage{graphicx}
\usepackage{xcolor}
\usepackage{amsmath}
\usepackage{subfigure}	
\usepackage{caption}
\usepackage{threeparttable}
\begin{document}

\title{Main Beam Modeling for Large Irregular Arrays
}
\subtitle{The SKA1-LOW telescope case}


\author{Ha Bui-Van \and
        Christophe~Craeye \and
        Eloy de Lera Acedo
}


\institute{Ha Bui-Van \at
              ICTEAM, Universit\'e catholique de Louvain, 1348 Louvain-la-Neuve, Belgium \\
              \email{buivanha@uclouvain.be}           
           \and
           Christophe Craeye \at
              ICTEAM, Universit\'e catholique de Louvain, 1348 Louvain-la-Neuve, Belgium \\
              \email{christophe.craeye@uclouvain.be}           
           \and
           Eloy de Lera Acedo \at
              Cavendish Astrophysics, Cavendish Laboratory, University of Cambridge \\
              \email{eloy@mrao.cam.ac.uk}           
}

\date{Received: 02/2017 / Accepted: date}

\maketitle

\begin{abstract}
Large radio telescopes in the $21^{\textrm{st}}$ century such as the Low-Frequency Array (LOFAR) or the Murchison Widefield Array (MWA) make use of phased aperture arrays of antennas to achieve superb survey speeds. The Square Kilometer Array low frequency instrument (SKA1-LOW) will consist of a collection of non-regular phased array systems. The prediction of the main beam of these arrays using a few coefficients is crucial for the calibration of the telescope. An effective approach to model the main beam and first few sidelobes for large non-regular arrays is presented. The approach exploits Zernike polynomials to represent the array pattern. Starting from the current defined on an equivalence plane located just above the array, the pattern is expressed as a sum of Fourier transforms of Zernike functions of different orders. The coefficients for Zernike polynomials are derived by two different means: least-squares and analytical approaches. The analysis shows that both approaches provide a similar performance for representing the main beam and first few sidelobes. Moreover, numerical results for different array configurations are provided, which demonstrate the performance of the proposed method, also for arrays with shapes far from circular.
\end{abstract}

\section{Introduction}
\label{intro}

Modern radar and astronomic systems often consist of phased aperture arrays composed of hundreds to thousands of complex antennas to provide the most sensitive observation. The low frequency instrument of the Square Kilometer Array (SKA) telescope~\cite{SKA_pp,SKA}, the SKA1-LOW telescope, covering from 50~MHz up to 350~MHz, will consist of 512 stations, each being made of 256 log-periodic antennas (SKALA)~\cite{Eloy15}. This will be the largest and most powerful radio telescope of the world at meter and centimeter wavelengths by the time of its completion in the early 20s. The calibration of such arrays is a challenging task, since the embedded element patterns (EEPs), i.e. the pattern of an antenna in the array when all other elements are terminated~\cite{Pozar94}, strongly varies among elements due to the effects of mutual coupling.
 
Generally, practical calibration, relying on the use of available (sky) reference sources, is time consuming and direction-dependent, which becomes difficult or impossible if the number of measurements is large. Recently, the prediction of array beam patterns has been proposed, providing a new perspective in calibration~\cite{Maaskant12,Davidson13}. A set of characteristic basis function patterns (CBFPs) is used in conjunction with a single far-field measurement to predict the array main beam and sidelobes. An average embedded pattern is calculated from these CBFPs with coefficients determined by least-square fitting the modeled covariance matrix of the output port to the  measured one. The array pattern is then the product of the common identical pattern and array factor.
Another effort to predict the array beam patterns has been presented in~\cite{Christophe11}, where the array factor has been represented using Zernike polynomials. Although in that work only the array factor was modeled, the main beam and few sidelobes are effectively modeled using a limited number of terms (i.e. coefficients).  

Thanks to the evolution in computational electromagnetics, simulation tools are now able to efficiently and accurately simulate the embedded element patterns of complex antennas in large arrays~\cite{DavidTAP11,Ludick14}, starting from integral-equation approaches, such as the Method-of-Moments, and an acceleration technique such as the Macro Basis Functions (MBF). The MBFs correspond to a limited set of basis functions for current distributions defined over a given antenna and described in terms of elementary basis functions~\cite{Mosig2000}. Implementing the interpolatory technique~\cite{DavidTAP11}, one is able to obtain the currents on all the antennas and all EEPs. This hence allows the fast computation of the array pattern for any array excitation law. Storing all the EEPs or the MBF expansion coefficients on each antenna may still be considered too expensive,  especially when a high resolution representation of patterns is needed. The spherical-wave expansion~\cite{Hansen88} is also considered to represent the embedded element patterns~\cite{EloyICEAA11},~\cite{HaICEAA17}; it can also be used to directly describe the array pattern over the whole hemisphere. However, the main problem with this representation is that the number of needed coefficients grows like the square of the array diameter in wavelengths~\cite{Hansen88},~\cite{Jensen04}. This remains true to a large extent for individual EEPs because of mutual coupling, which often involves the whole array; this problem besides the fact that there are as many EEPs as elements.

For some applications such as the SKA, in most circumstances, a limited angular field of view is of interest, such that a compact representation of the main beam and first few sidelobes is sufficient. Furthermore, the windowing effect provided by the interferometric operation weights down the effect of the far out sidelobes, stressing the particular relevance of mapping the main beam and first few sidelobes. This paper addresses a possible solution by representing the array patterns using Hankel transforms of Zernike polynomials, where the main beam and first few sidelobes are accurately modeled using a limited number of coefficients. 
Spherical-wave expansions do not benefit (in terms of required order) from a restriction of the field of view, since they inherently aim at the representation of patterns over the whole hemisphere (strictly speaking over the whole sphere).
The method based on Zernike functions, as presented here, is an extension of the work in~\cite{Christophe11} to fully coupled arrays. More precisely, the array patterns are accurately calculated using a fast simulation technique, which is then effectively represented using a series involving Zernike functions. The proposed method differs from~\cite{Christophe11} in that it works directly at the array pattern level instead of the array factor level, therefore taking into account the mutual coupling in the array. Moreover, by working at the array pattern level, the method requires much fewer coefficients than the work in~\cite{Christophe11}. Assuming the work in~\cite{Christophe11} can be extended to the array pattern level, (i.e. the superposition of the products between array factors and MBF patterns~\cite{Christophe09}) and denoting as $N_{mbf}$ the number of MBF, it will require approximately $N_{mbf}/2$ times more coefficients to model the full array pattern,  as compared with the approach proposed in this paper. 
It is also worth noting that despite similarities between the proposed method and aperture theory, there are several differences that make the present method more general; and these points will be highlighted throughout the paper. 

Arrays made of SKALA elements are considered in this study. The interpolatory technique~\cite{DavidTAP11}, renamed as HARP and extended in ~\cite{QuentinAPS15}, \cite{Ha17}, is implemented to analyze the arrays and to calculate the EEPs. The main beam and first few sidelobes of these arrays are then modeled using the proposed technique, including the SKA station and other different array configurations. The author would like to make it clear that this approach is not efficient for the representation of far-out sidelobes.

The remainder of this paper is organized as follows. In Section 2, the beam modeling technique is presented. Starting from the currents, defined on an aperture just above the array, we then derive the representation of the pattern as a sum of Fourier Transforms (FT) of different Zernike polynomial orders. The coefficients are determined in two different ways, namely the least-squares and analytical approaches. Several useful properties of the Fourier transform  are exploited to adapt the approach to different array configurations, including strongly excentric ones. The fast full-wave simulation technique to analyze the considered arrays is presented in Section 3. In Section 4, numerical results for beam modeling for different array configurations are presented. The results are analyzed for different combinations of Zernike polynomial orders. The paper ends with a discussion.

\section{Beam Modeling Technique}
\label{sec:1}

In this section, a technique for obtaining a compact representation of the main beam and first few sidelobes of different non-regular arrays is detailed. This involves the link between the radiation of current distributed over an aperture and the decomposition of those currents into Zernike functions. The coefficients multiplying the Zernike basis functions are then derived by different means.
In the following, $I_x$ denotes  the $x-$component of a ``pseudo-current'', whose Fourier transform simply corresponds to the $x-$component of the pattern $G_x(\phi,\theta)$ (following the Ludwig's first definition~\cite{Ludwig73}).  For more detail about the radiation from aperture currents, the reader is referred to~\cite[Chapter 7]{Kildal15}, where radiated fields in spectral Cartesian coordinates and radiation from circular apertures are formulated in Section 7.3 and 7.5, respectively. The link between ``pseudo-currents'' and ``equivalent currents'' defined in~\cite{Poggio73},~\cite{Harrington01} is detailed in the Appendix I. 

\subsection{Decomposition of Aperture Distribution}
\label{sec:2}

We consider an infinite plane $S$ located in $x-y$ Cartesian coordinates just above the radiating structure with a normal unit vector $\hat{n}$ pointing away from $S$. A given component of the aperture pseudo-current distribution, $I_x(r,\alpha)$, will be described in polar coordinates; it is assumed to have non-zero values within a domain limited to the disk $r<b$. A Fourier-series decomposition yields:
\begin{equation}
	I_{{x}}(r,\alpha) = \sum_{n=-N}^N{a_{nx}(r)~e^{jn\alpha}} \label{eq:FTS}
\end{equation}
with
\begin{equation}
	a_{nx}(r) = \frac{1}{2\pi}~\int_{0}^{2\pi}{I_x(r,\alpha)~e^{-jn\alpha}~d\alpha}
\end{equation}

The distribution can be decomposed in terms of circle polynomials~\cite{Savov03}. It appears that Zernike was the first to propose this type of decomposition~\cite{Zernike34}. Hereafter, we will hence name this the Zernike decomposition. The completeness and orthogonality of Zernike polynomials over the unit disk allows the following decomposition:
\begin{equation}
 a_{nx}(r) \simeq \sum_{m=0}^{M}{z_{mnx}F_m^{\mid n\mid}(r/b)}	\label{eq:ZernikeRep}
\end{equation}
with
\begin{equation}
	z_{mnx} = \frac{({\mid}n{\mid} + 2m +1)}{\pi b^2}B_{mnx}
\end{equation}
and
\begin{align}
	B_{mnx} &= 2\pi \int_{0}^{b}{F_m^{\mid n\mid}(r/b) \,a_{nx}(r)\, rdr} \\
	&= \int_{S}{I_x(r,\alpha) \, F_m^{{\mid}n{\mid}} (r/b) \, e^{-jn\alpha} dS} \label{eq:Bmn}
\end{align}
where $F_m^{{\mid}n\mid}(\rho)$ is the modified circle polynomial of order $m$ and degree $\mid$${n}$$\mid$~\cite{Savov03}.

\subsection{Radiation Integral Representation} \label{subsec:representation}
The $x-$component of the radiation pattern can be associated to a pseudo-current distribution $I_x(r,\alpha)$ through the following inverse Fourier transform:
\begin{equation}
	G_x(\phi,\theta) = \int_S{I_x(r,\alpha) \, e^{jk(u_{x}x + u_{y}y)} dS} \label{eq:pt1}
\end{equation}
where $k$ is the wavenumber, $\hat{u} \equiv (u_x,u_y,u_z)$ is a unit vector pointing in the direction of observation.  The equation can be re-written in cylindrical coordinates:
\begin{align}
	G_x(\phi,\theta) &= \int_0^b \int_0^{2\pi} I_x(r,\alpha)~e^{jkr \sin \theta \cos (\phi-\alpha)} d\alpha~r dr\\
	&= \sum_{n=-N}^Nj^n e^{jn\phi}  {\int_0^b{a_{nx}(r) 2\pi  J_n(kr\sin \theta) rdr}} \label{eq:pattern}
\end{align}
where the last equation has been obtained with the help of Eq.~(\ref{eq:FTS}) and the integral representation of Bessel functions. Introducing into Eq.~(\ref{eq:pattern}) the order$-n$ Zernike polynomial representation, Eq.~(\ref{eq:ZernikeRep}), the following expressions are obtained:
\begin{align}
	G_x(\phi,\theta) &\simeq \sum_{n=-N}^N j^n e^{jn\phi} \sum_{m=0}^{M} 2\pi  \int_0^b z_{mnx}F_m^{\mid n\mid}(r/b) J_n(kr\sin \theta) r dr\\
	&= \sum_{n=-N}^N j^n e^{jn\phi} \sum_{m=0}^{M} ({\mid}n{\mid} + 2m + 1) (-1)^s\frac{J_{\mid n\mid + 2m + 1}(kb\sin \theta)}{kb\sin \theta} B_{mnx} \label{eq:FTZ}
\end{align}
with $s=0$ if $n\geq 0$ and $s=n$ if $n < 0$. Eq.~(\ref{eq:FTZ}) is obtained with the help of the Hankel transform of the Zernike polynomials~\cite{Savov03}. This representation can be linked with representations of field radiated by apertures, in which Jacobi functions are exploited~\cite{RahmatSamii80}. 
The aperture theory, however, generally considers a flat 2-D surface with known current/field distribution. In this paper, we assume that the pattern in~(\ref{eq:pt1}) is known in advance, i.e. via simulation or measurement, the coefficients multiplying the Zernike polynomials are then derived using relation in~(\ref{eq:Bmn}) or~(\ref{eq:FTZ}) following the analytical or the least-squares approaches presented in the next subsection, respectively. This allows the proposed technique to be applicable to arrays of antennas, including arrays made of 3-D antennas, as presented in this paper.
Moreover, while the aperture theory works with continuous current distribution on the surface, the proposed method is applicable for arrays of connected or disconnected antennas with arbitrary shapes and configurations, including random or irregular antenna distributions.

Eq.~(\ref{eq:FTZ}) is a compact representation of the pattern in Eq.~(\ref{eq:pt1}) using Hankel transforms of the Zernike polynomials with corresponding coefficients.  In the following section, these coefficients are determined using either analytical or in least-squares approaches.

\subsection{Calculation of Coefficients} \label{subsec:coeff}

It is clear that, if the current distribution on the surface is available, the coefficients are readily obtained by projecting the current $I_x(r,\alpha)$  on the set of basis functions following expression~(\ref{eq:Bmn}). From the beam modeling point of view, the array pattern is assumed to be available, e.g. from simulations, the coefficients are then computed following two approaches presented in this section.

\subsubsection{Least-Squares Approach} \label{sec:LSA}

From Eq.~(\ref{eq:FTZ}), the Hankel transform of Zernike polynomial is available for given values of $m$, $n$ and $\rm{K} = kb\sin\theta$. The equation can be reformatted as follows:
\begin{equation}
	\rm{p} = H\rm{c}
\end{equation}
where $\rm{p}$ is a column vector describing the pattern that needs to be modeled (one entry per direction), H is the matrix containing the Hankel transform of Zernike polynomial of order $m$ and degree $n$ (each column corresponds to a pair of orders ($m,n$)), and $\rm{c}$ is a vector of coefficients. The latter is efficiently calculated in the least-squares sense.

\subsubsection{Analytical Approach} \label{sec:AA}

The second approach exploits the link between the current and aperture field as presented above. The current is obtained from the Fourier transforms of the pattern in Eq.~(\ref{eq:pt1})  as follows:
\begin{equation}
	I_x(x,y) = \frac{1}{(2\pi)^2}\iint_{k_x,k_y} G_x(\phi,\theta)\,e^{-j(k_x x + k_y y)} \,dk_x\,dk_y	 \label{eq:current}
\end{equation}
where $k_x=k\sin \theta\cos\phi$ and $k_y = k\sin\theta \sin\phi$. If $G_x(\phi,\theta)$ is calculated or measured on a regular ($k_x,k_y$) grid, the evaluation of Eq.~(\ref{eq:current}) is accelerated by 2D Fast-Fourier transform (FFT). Once the current is attained, the coefficients are calculated through Eq.~(\ref{eq:Bmn}). 

Eq.~(\ref{eq:current}) is integrated over the visible region, i.e. $k_x^2 + k_y^2 \leq k^2$, and the obtained current corresponds to the currents radiating into the visible space. 
The current $I_x$ in~(\ref{eq:current}) may not be the actual ``physical'' current on the surface, as the pattern in~(\ref{eq:pt1}) can be obtained from an array of 3-D antennas. It is rather a ``pseudo-current'' that produces the pattern $G_x$ and is an intermediate quantity allowing one to obtain the coefficients of Zernike polynomials via~(\ref{eq:Bmn}). These coefficients are then exploited to model the pattern using~(\ref{eq:FTZ}).
 
\begin{figure}[!htb]
\centering 
	\includegraphics[scale=0.65,clip,trim={0cm 0cm 1cm 0cm}]{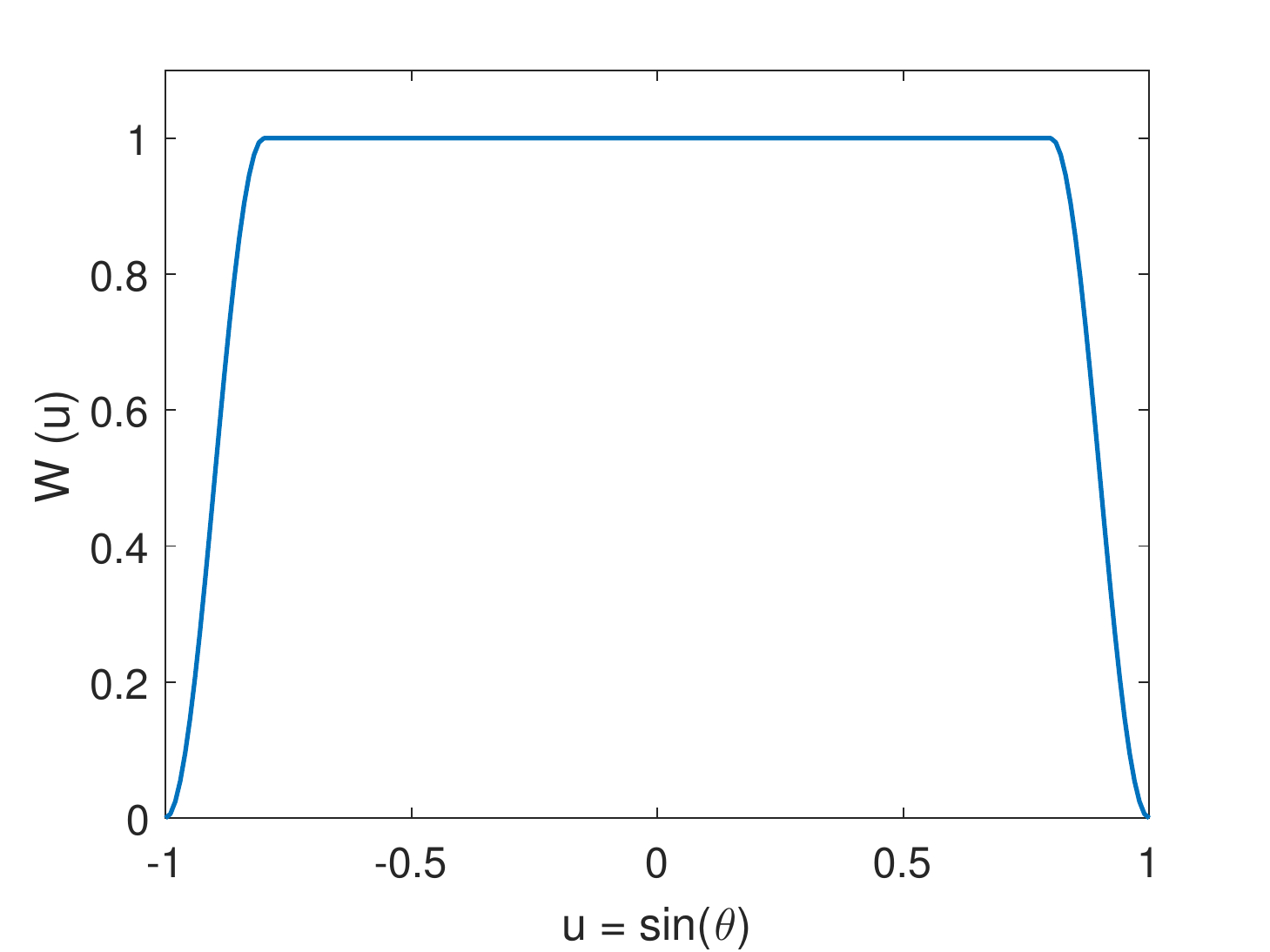}
	\caption{Windowing function with Sigmoid-like-function tails used to reduce the Gibbs phenomenon.} \label{fig:windown}
\end{figure}

In order to reduce the ripples of the current in the space domain, known as the ringing artifact or the Gibbs phenomenon~\cite{Hewitt79}, a windowing function with a Sigmoid-like tail, shown in Fig.~\ref{fig:windown}, is multiplied to the pattern before the FT operation. 
The weighting function is defined as follows:
\[ \rm{w}(u) =
  \begin{cases}
    1       & |u| \leq 1 - 2d  \\
    \frac{1}{2} - \frac{1}{2}\sin (\frac{(u-1+d)\pi}{2d})    &  1-2d \leq |u| \leq 1\\
  \end{cases}
\]
where $d$ is the half-width of the tail. The weight, $\rm{w}$, is multiplied to the pattern before performing Fourier Transforms, which will scale down the  pattern in certain direction. The sacrificed region is mainly at far-out sidelobes or at grazing angles, which will not impact on the quality of the modeled pattern as the power level in that region is relatively low compared to the main beam, e.g. below -30\,dB. The main-beam and first sidelobes are unaffected, since they are multiplied by 1. 
In these examples, the value of $d$ is set to $0.01$. This value is selected such that the tail would not cover the region of interest. The $d=0.01$ choice leads to patterns being unaffected up to $78^\circ$ from zenith, which is larger than the maximum angle of observation currently foreseen for the SKA1-LOW telescope~\cite{SKAReq}. For other systems, smaller values of $d$ can be used to address larger scanning angles without dramatic impact on accuracy.
 Performing this step results in a smooth current in Eq.~(\ref{eq:ZernikeRep}), and subsequently the reduction of the number of terms (orders) needed to represent that current. Moreover, since the  aperture or the antenna array is finite, the current, calculated in Eq.~(\ref{eq:current}), is truncated just to the side of the aperture adding one wavelength in all directions. The coefficients $B_{mnx}$ are now readily obtained using Eq.~(\ref{eq:Bmn}).

So far, the pattern representation is discussed for only one component of the current, i.e. $I_x$. At first glance, it appears that the number of coefficients to be used is equal for the other component, i.e. $I_y$. The total coefficients then will be doubled when  the polarimetric pattern description is required, as opposed to the scalar (e.g. power-based) case.  This is not exactly true. Suppose $X$ is the co-polar component, then the cross-polar one, which is supposed to be significantly smaller, would require fewer terms to achieve a comparable level of error, defined in term of residual current. The examples in Section~\ref{sec:beamModeling} will illustrate that only half of the coefficients is required for the cross-polarization to achieve an absolute error similar to that of the co-polarization pattern. The full pattern then can be easily obtained from $G_x(\phi,\theta)$ and $G_y(\phi,\theta)$. Other polarization representations can be also derived through the polarization matrix transform presented in~\cite{Ludwig73}.

\subsection{Array Configurations}  \label{subsec:arrayconfig}
In the proposed method, the aperture current and the pattern are linked through the Fourier Transforms (FT) in an analytical form such that we can exploit different properties of FT to facilitate the modeling of different aperture/array configurations. Exploiting this advantage, scanned arrays and arrays with excentric shapes can be addressed as well. 

\subsubsection{Scanned Beam Patterns} \label{subsubsec:scannedarray}

For the scanned beam, the array pattern is calculated as:
\begin{equation}
G_x(\theta,\phi) = \sum_{n=1}^{\rm{NA}} e_{n}(\theta,\phi)\,w_n\,e^{-j(u_{xo}x_n + u_{y0}y_n)} \label{eq:sArray}
\end{equation}
where $e_{n}(\theta,\phi)$ is the EEP of the  $n^{th}$ antenna, NA is the total number of antennas, $w_n$ is the weight of the excitation vector, $(x_n,y_n)$ is the position of antennas, $(u_{xo}, u_{yo})=(k\sin{\theta}_0\cos{\phi}_0,k\sin{\theta}_0\sin{\phi}_0)$ and (${\theta}_0,{\phi}_0$) is the scanned direction i.e. angle from broadside and azimuth, respectively.
The equivalent current obtained from the Fourier Transforms of the pattern using (\ref{eq:current})  on the aperture will exhibit an approximately linear phase progression. This will complicate the modeling process, as the number of required coefficients will rapidly increase. One way to overcome this difficulty is to first un-wrap the phase shift required to scan the beam using~(\ref{eq:unwrapPhase}), then to model the current $I_x^u$ (using Eq.~(\ref{eq:ZernikeRep}) and Eq.~(\ref{eq:Bmn})\,), and finally to shift the pattern in opposite direction to re-introduce the phase gradient, thanks to the shifting property of the Fourier Transform:
\begin{equation}
I_x(x,y) = I_x^{u}\,e^{j(u_{xo}x + u_{yo}y})   \label{eq:unwrapPhase}
\end{equation}
This is another point about which the proposed method differs from aperture theory. For scanned arrays, the phase-shift is applied for each antenna as in~(\ref{eq:sArray}), instead of a phase progression on the 2-D surface in aperture theory. This makes the proposed technique  more general and applicable to arrays of different types of antennas.

\subsubsection{Aperture Shapes (Array Configurations)} \label{sec:AS}
It is true that due to their properties, the Zernike polynomials (or circle polynomials) are best fit to decompose fields in circular apertures. However, the proposed method is not limited to these shapes, as will be discussed in Section~\ref{sec:beamModeling} for different array configurations, e.g. ellipse, hexagon, rectangular and pentagon shapes. 
The main beam and first few sidelobes are successfully modeled using the proposed method, regardless of the aperture shapes.
Moreover, one can also benefit from the scaling property of the Fourier Transform to deal with apertures with strongly excentric shapes. For example, to model the pattern of current on a flat elliptical aperture, one may scale the current in space domain to a circular distribution. In the spectral domain, for the radiation pattern, one just needs to scale down by the inverse factor in the same direction. As a result, the method performs equally well as in the case of a circular aperture, and the modeled beam essentially requires the same number of coefficients as for a nearly circular array. 

\section{Case Study: Arrays of Log-periodic Antennas}\label{sec:LPDAs}

\begin{figure}[!htb]
\centering
	\includegraphics[scale=0.8,clip,trim={1cm 0cm 1cm 2.5cm}]{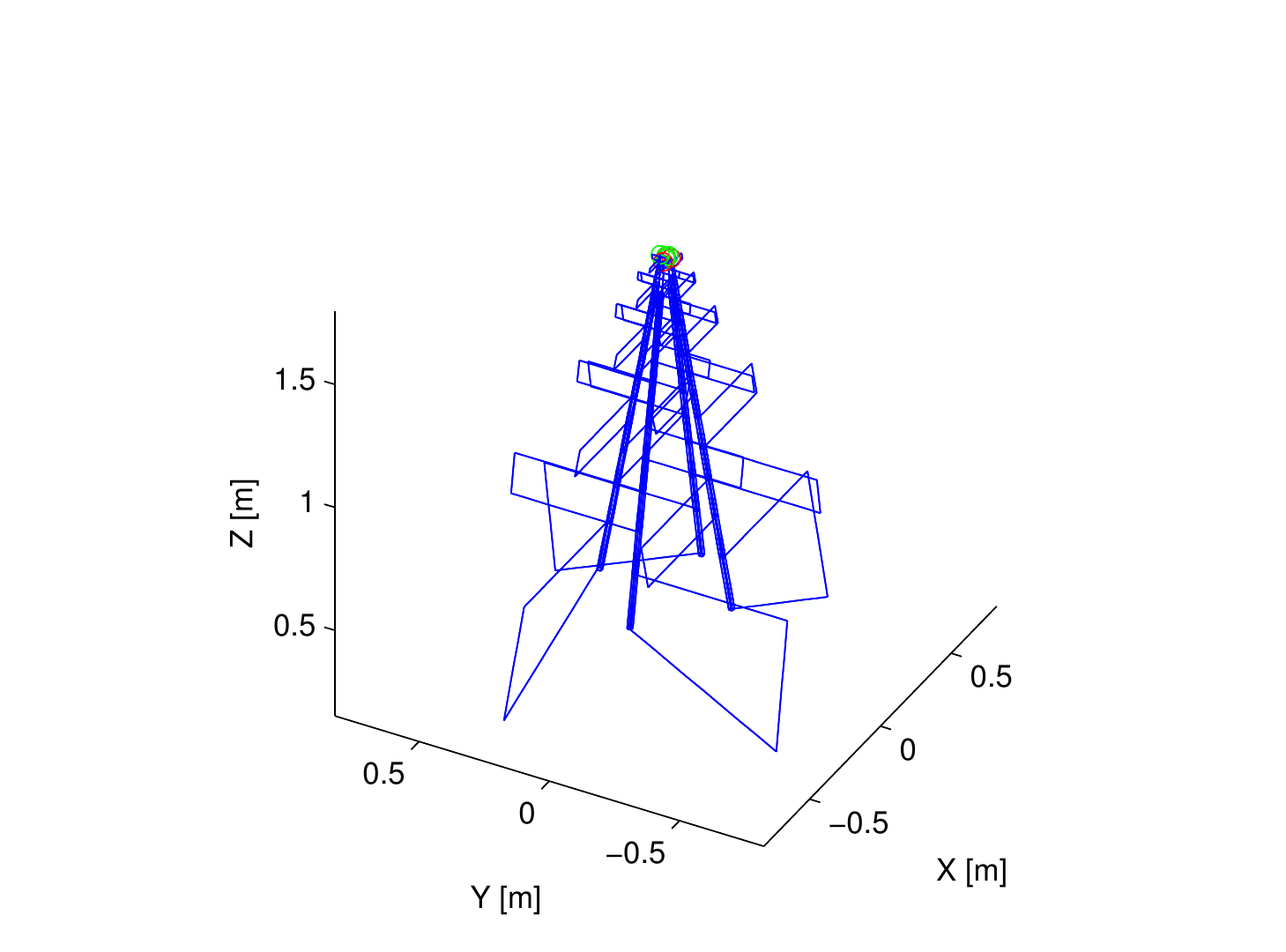}
	\caption{3D-view of the SKA Log-Periodic Antenna. The antenna consists of 2$\times$2 arms, allowing dual-polarized observation.} \label{fig:meshSKALA}
\end{figure}

Arrays consisting of the log-periodic antennas (SKALA)~\cite{Eloy15}, designed for the SKA1-LOW array, are considered  in this paper. The SKALA has a footprint of 1.2$\times$1.2\,$\rm{m}^2$  and it is 1.8\,m in height, as plotted in Fig.~\ref{fig:meshSKALA}. The chosen element represents a general and complex structure with practical applications in the field of radio astronomy~\cite{Eloy15}. It is important to note that the above proposed method is straightforward and is applicable to model the beam pattern of any array, regardless of the type of element. 

A thin-wire mesh is used to represent the SKALA, except for the thick spine support, which has been modeled as a cage made of thin wires~\cite{QuentinAPS15}. A $100~\rm{\Omega}$ loading impedance is attached to the two basis functions that cross the feeding ports of the dual-polarized SKALA.  
All arrays made of SKALA elements are analyzed using the HARP method~\cite{QuentinAPS15}. The EEPs and array patterns are obtained, from which a compact representation of the main beam and first few sidelobes are derived using the proposed approach as will be shown in the next section. Readers interested in the HARP method are referred to~\cite{DavidTAP11,QuentinAPS15,Ha17} for a detailed description of the method.

\section{Numerical Results}  \label{sec:beamModeling}

Numerical results for beam modeling for different array configurations are presented in this section. HARP is called first to analyze the array and to calculate all array patterns, which are then modeled using the proposed method. In our study, the patterns are represented by three components ${G_x}$, ${G_y}$, and ${G_z}$, corresponding to the Ludwig's first definition~\cite{Ludwig73}. The proposed method is then applied to model ${G_x}$, ${G_y}$, while ${G_z}$ is automatically obtained from ${G_x}$, ${G_y}$. 
 It is worth noting that, for the analytical approach, ${G_x}$ and ${G_y}$ are exploited to obtain the ``pseudo-currents'' (Section~\ref{subsec:representation}) over a surface on top of the array  via  the inverse Fourier transform~(\ref{eq:current}). However, the ``pseudo current''  does not correspond to the physical current, nor to the ``equivalent current'' often used in electromagnetics~\cite{Harrington01}. The relation between the ``pseudo current'' and the equivalent current for the aperture distribution is presented in Appendix I.
The SKALA is a dual polarized antenna, and in this study, elements are excited such that ${G_x}$, and $\rm{G_y}$ are the co-polar and cross-polar components, respectively. For the other excitation, the same performance is achieved, which will not be discussed. We qualify the modeling by showing the exact pattern (${G_x}$), the represented pattern (${G_x\_}_{app}$), and the difference (error) by using least-squares (LS) and analytical approaches (see Section~\ref{subsec:coeff}). The error is measured by the subtraction of complex values and is carried out before taking the magnitude of the result, followed by averaging over $\phi$ for all these examples. All the arrays analyzed below are simulated at 110\,MHz using 20 Macro Basis Functions (MBFs).

\begin{figure}[!htb]
\centering
	\includegraphics[scale=0.65,clip,trim={0cm 0cm 1.5cm 0cm}]{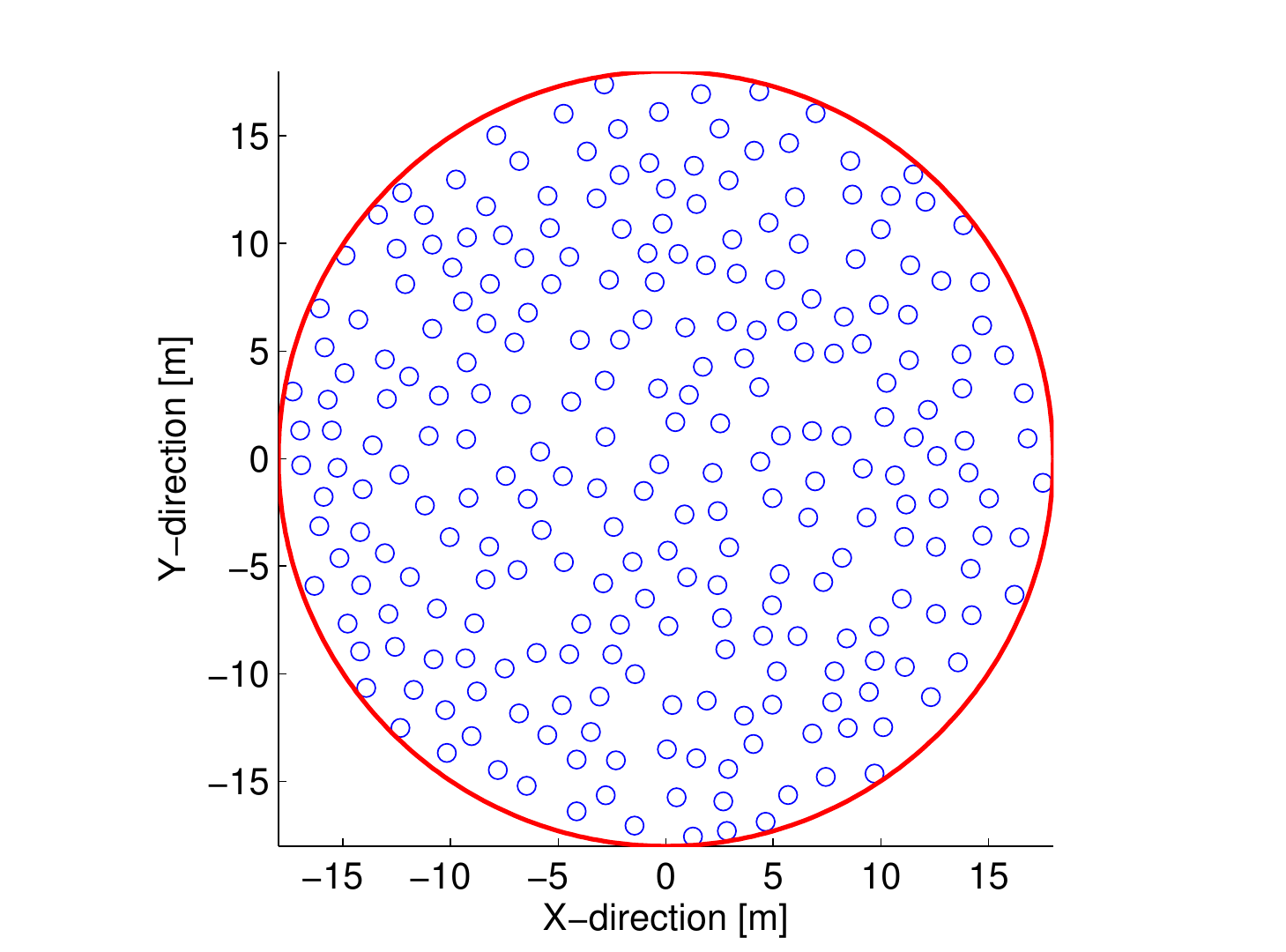}
	\caption{Configuration of a SKA station with 256 SKALA (each circle is an antenna in Fig. ~\ref{fig:meshSKALA}). The layout resembles a random array over a circle of 35\,m diameter, with minimum distance between antennas of 1.37\,m.} \label{fig:cArray}
\end{figure}

\begin{figure}[!htb]
\centering
\subfigure[Array scanned at broadside]{	\includegraphics[scale=0.31,clip,trim={0.5cm 0cm 1cm 0cm}]{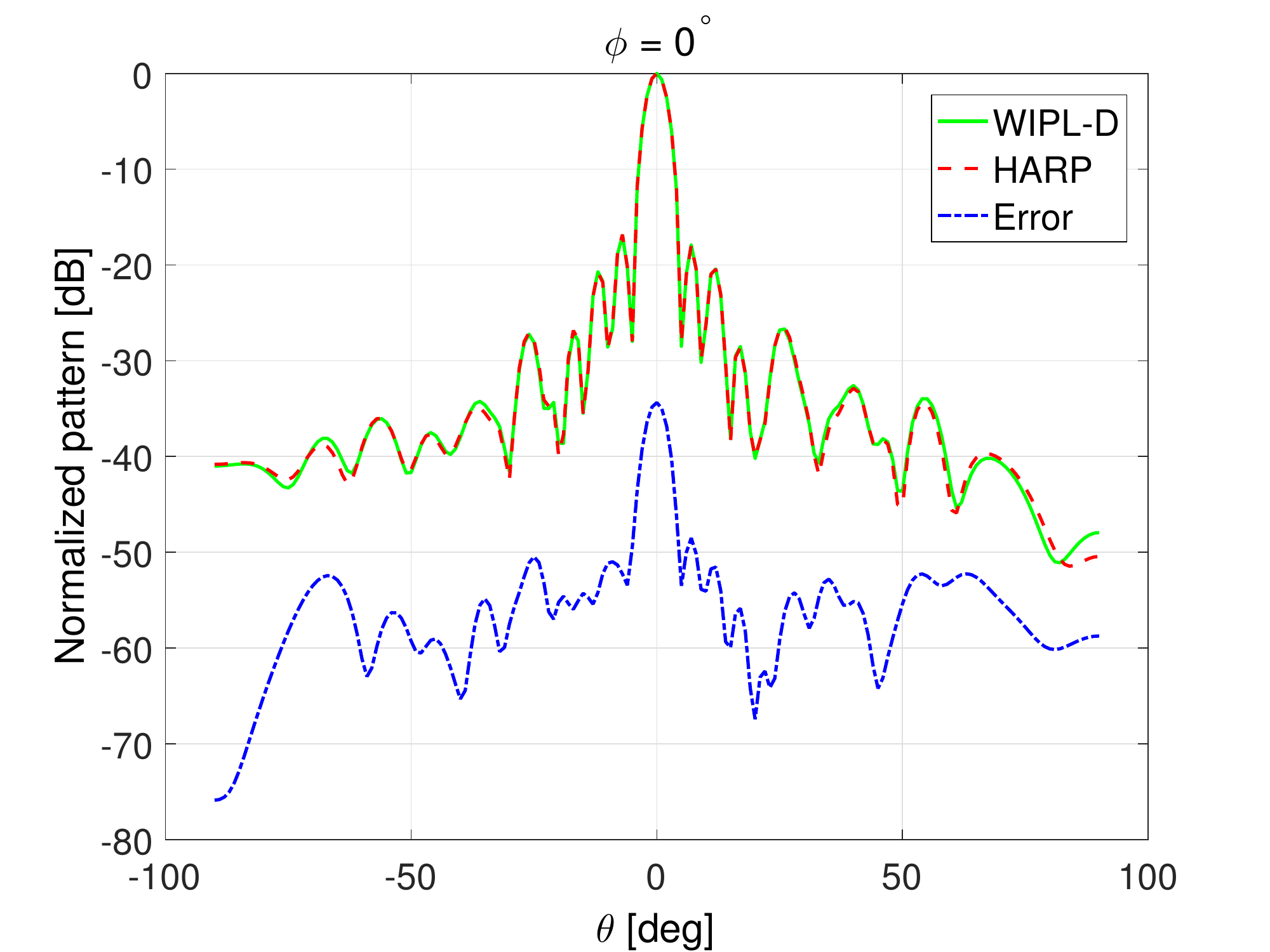}}
\hspace*{1mm}
\subfigure[Array scanned at $(\phi,\theta) = (0^\circ,30^\circ)$]{\includegraphics[scale=0.42,clip,trim={0.5cm 0cm 1cm 0cm}]{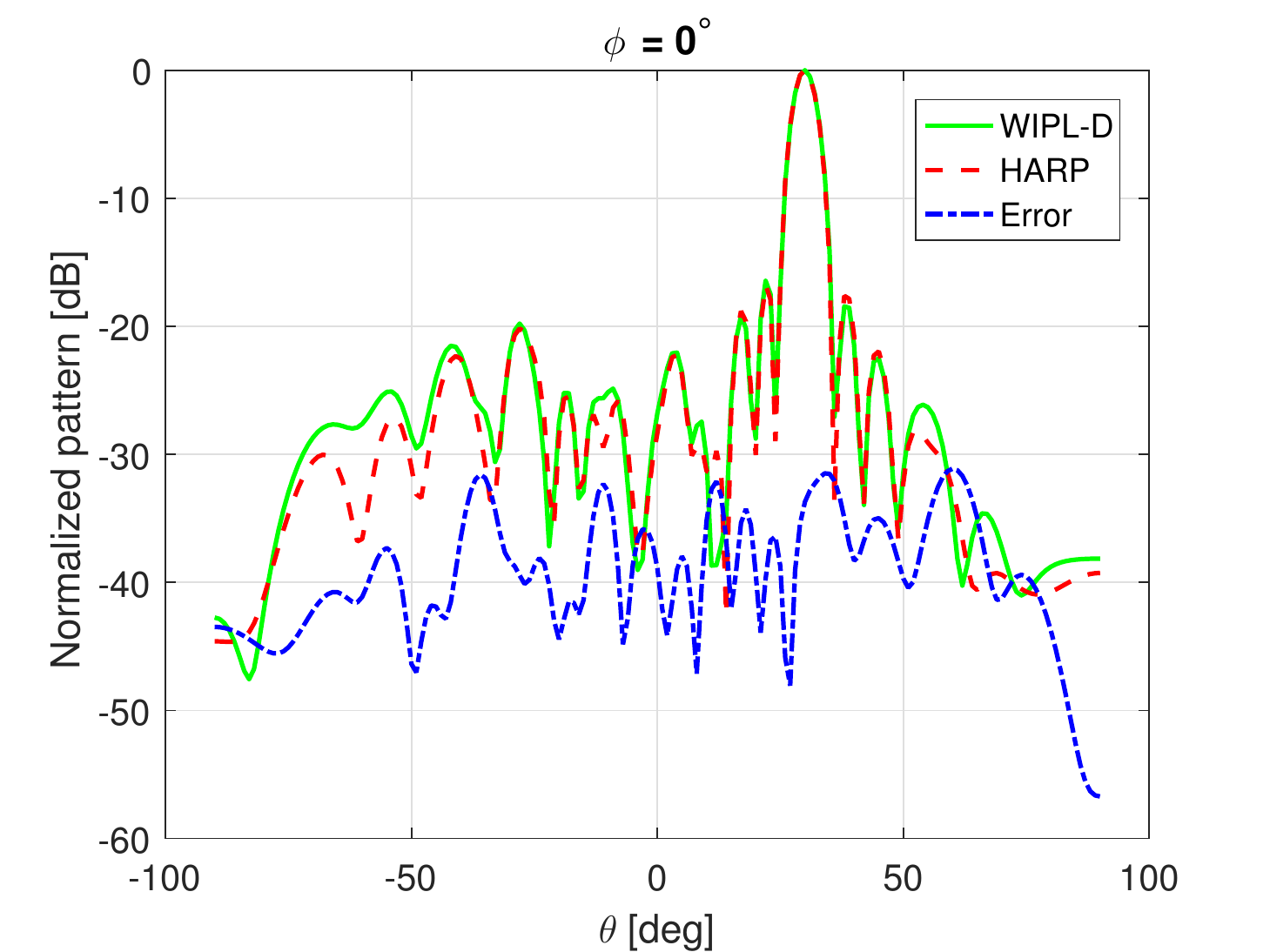}}
	\caption{Array pattern of uniform-excited SKA station at 110\,MHz, $\phi = 0^\circ$ cut: (a) scanned at broadside and (b) scanned at $(\phi,\theta) = (0^\circ,30^\circ)$. The patterns are calculated using a commercial software, WIPL-D (green line), the HARP (red dashed line). The difference (error) between HARP and WIPL-D are shown as well (blue dotted-dashed line). } \label{fig:cArray_validateWIPLD}
\end{figure}

\subsection{SKALA Array} \label{subsec:cArray}

The firstly considered example is the station of SKA1-LOW, consisting of 256 SKALA antennas, as shown in Fig.~\ref{fig:cArray}. The antennas are randomly distributed on a circular surface with a radius of 17.5\,m and a minimum distance between antennas of 1.37\,m. The array is analyzed using HARP; the embedded patterns and the array patterns are then calculated. Fig.~\ref{fig:cArray_validateWIPLD} plots the array pattern of an uniformly excited SKA1-LOW station simulated using the WIPL-D commercial software~\cite{wipld} and HARP. The results show an excellent agreement between two software even when the array is scanned far from broadside, which confirms the accuracy of HARP. From now on, the pattern obtained using HARP will be exploited  to derive the coefficients for Zernike polynomials following the proposed approach. It is important to note that to obtain all the element patterns of the SKA station, only a few minutes are required using HARP, while it took several days by using WIPL-D.

\begin{figure}
\centering
\subfigure[M = 3, N = 3]{\includegraphics[scale=0.375,clip,trim={0.25cm 0cm 1cm 0cm}]{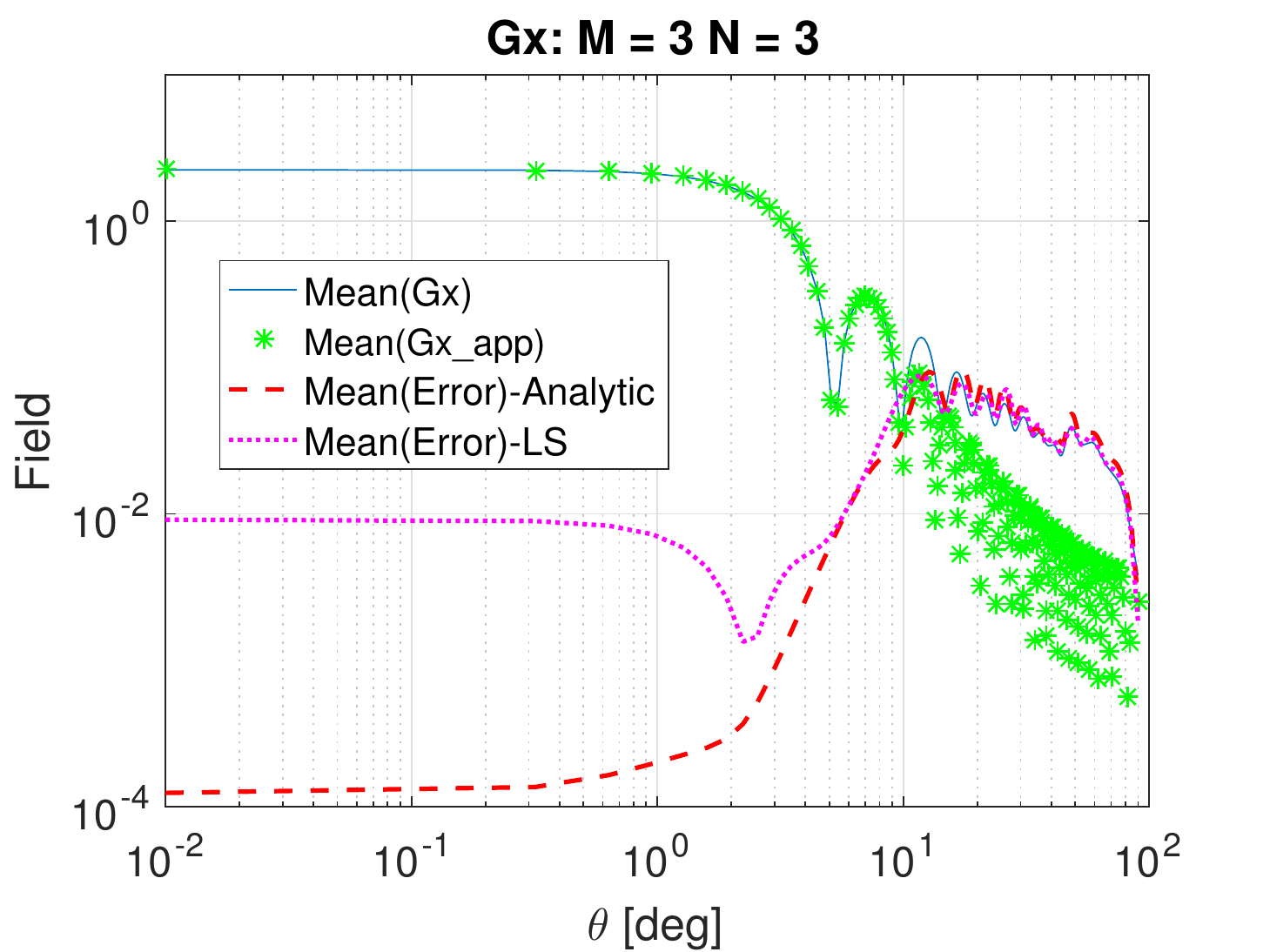}}
\hspace*{1mm}
\subfigure[M = 4, N = 4]{\includegraphics[scale=0.375,clip,trim={0.25cm 0cm 1cm 0cm}]{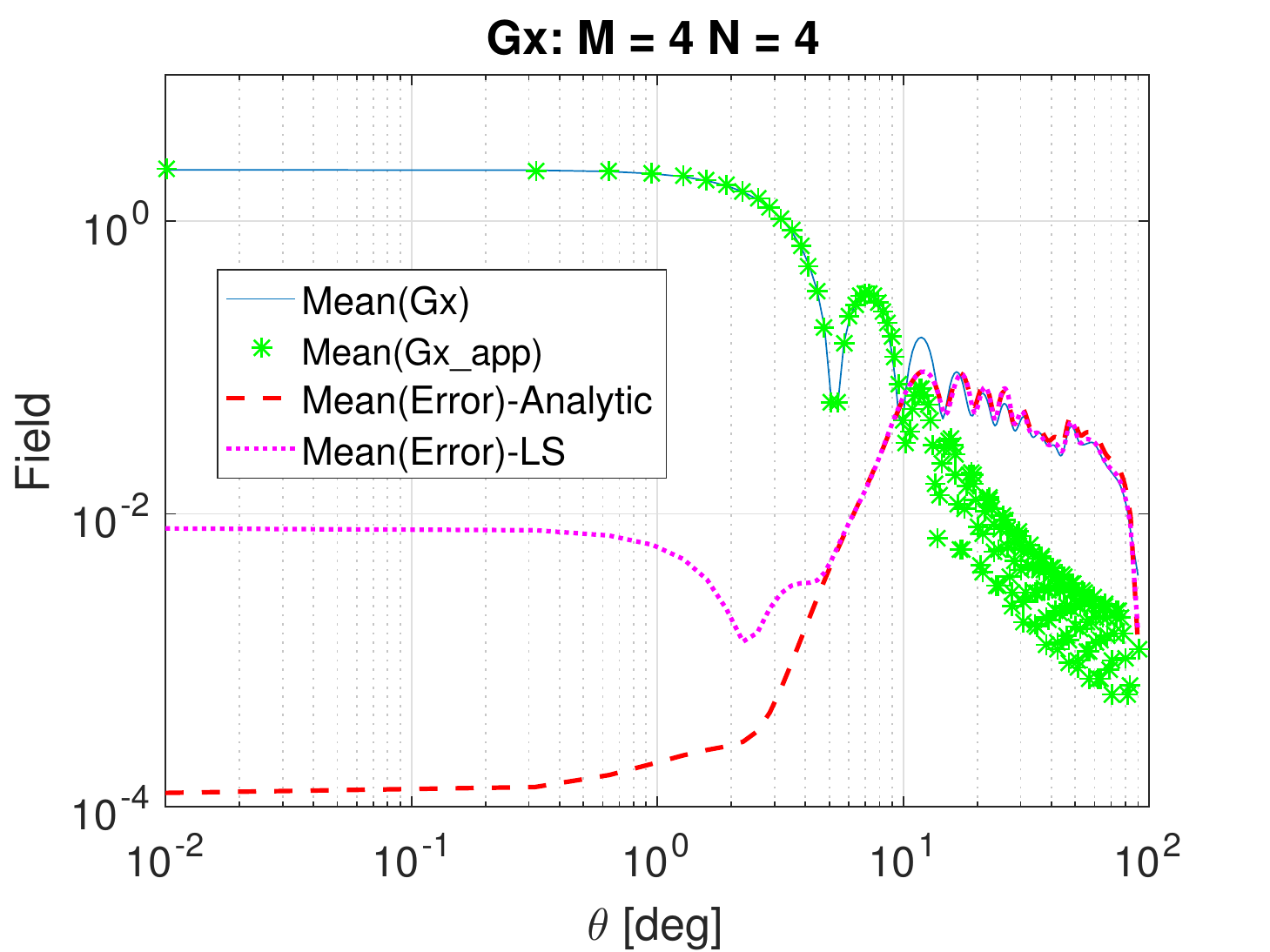}}
\vspace*{1mm}
\subfigure[M = 5, N = 5]{\includegraphics[scale=0.375,clip,trim={0.25cm 0cm 1cm 0cm}]{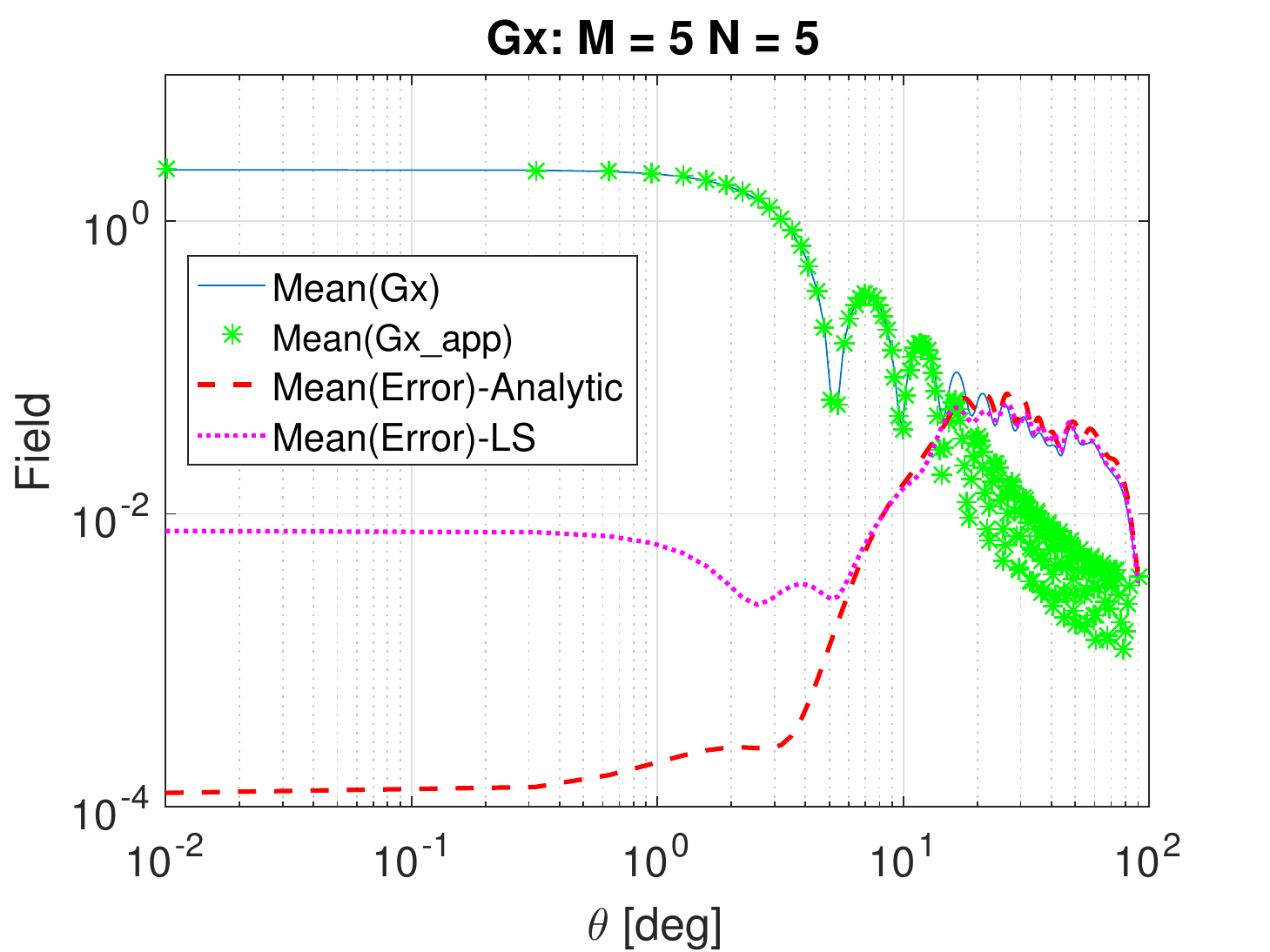}}
\hspace*{1mm}
\subfigure[M = 7, N = 7]{\includegraphics[scale=0.375,clip,trim={0.25cm 0cm 1cm 0cm}]{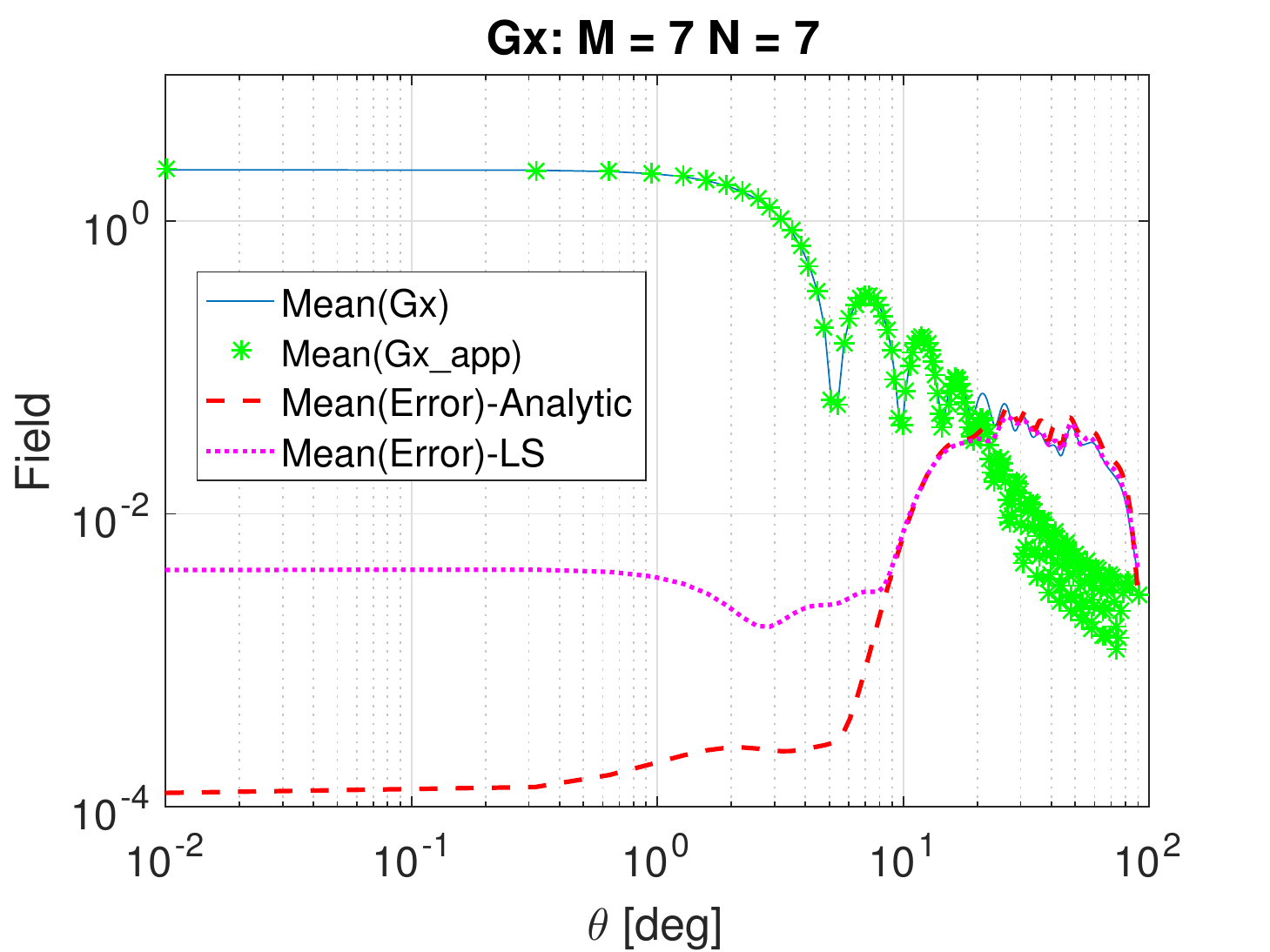}}
	
	\caption{Mean of ${G_x}-$pattern over $\phi$ of the SKA1-LOW station for different values of (M, N). The blue line is exact pattern calculated using HARP, green-star is the modeled pattern using the proposed method, and dashed-line is the error of the modeled pattern (using analytical approach (red) and least-square approach (pink)).} \label{fig:px_cArray}

\end{figure}

The co-polarization component of the pattern, ${G_x}$, is first modeled using the proposed method. The mean value of ${G_x}$ over $\phi$ is plotted as shown in Fig.~\ref{fig:px_cArray}. It is seen that the accuracy of modeled beam increases as the orders of Zernike polynomials grows. Interestingly, the coefficients, derived by both the least-squares and analytical approaches, exhibit a similar performance in the sidelobe area, the difference at the main beam being negligible. 
The typical achieved level of error will be illustrated below.
First, one may consider the maximum error w.r.t. the maximum of the main beam within the region extending up to the maximum of the first sidelobe. For a $1\%$ error, the method needs only 28 coefficients, i.e. $M=3$ and $N=3$, which is much smaller than in approaches devoting  only one calibration coefficient per antenna.
Next, in~\cite{SKAReq2}, the requirements are provided for the beam modeling of SKA1-LOW antennas. There, within the -6\,dB level, a $0.02\%$ root-mean-square error is requested at 110\,MHz. With the $M=4$ and $N=4$ parameters, the error level at -6\,dB is 0.011$\%$. For higher accuracy, i.e. modeling further in the sidelobe area, one just needs to increase the order of the Zernike polynomials. As shown in Fig.~\ref{fig:px_cArray},  66 ($M=5$ and $N=5$) and 120 ($M=7$ and $N=7$) coefficients are sufficient to properly model the second and third sidelobes, respectively. However, if one wants to include all the sidelobes, one might need more coefficients than the total number of antennas. In this case, it might be easier to use the EPPs and their coefficients.  The results indicate that the proposed method is very effective for applications in which only the main beam and the first few sidelobes are needed.

\begin{figure}[!htb]
\centering
	\includegraphics[scale=0.65,clip,trim={0cm 0cm 0cm 0cm}]{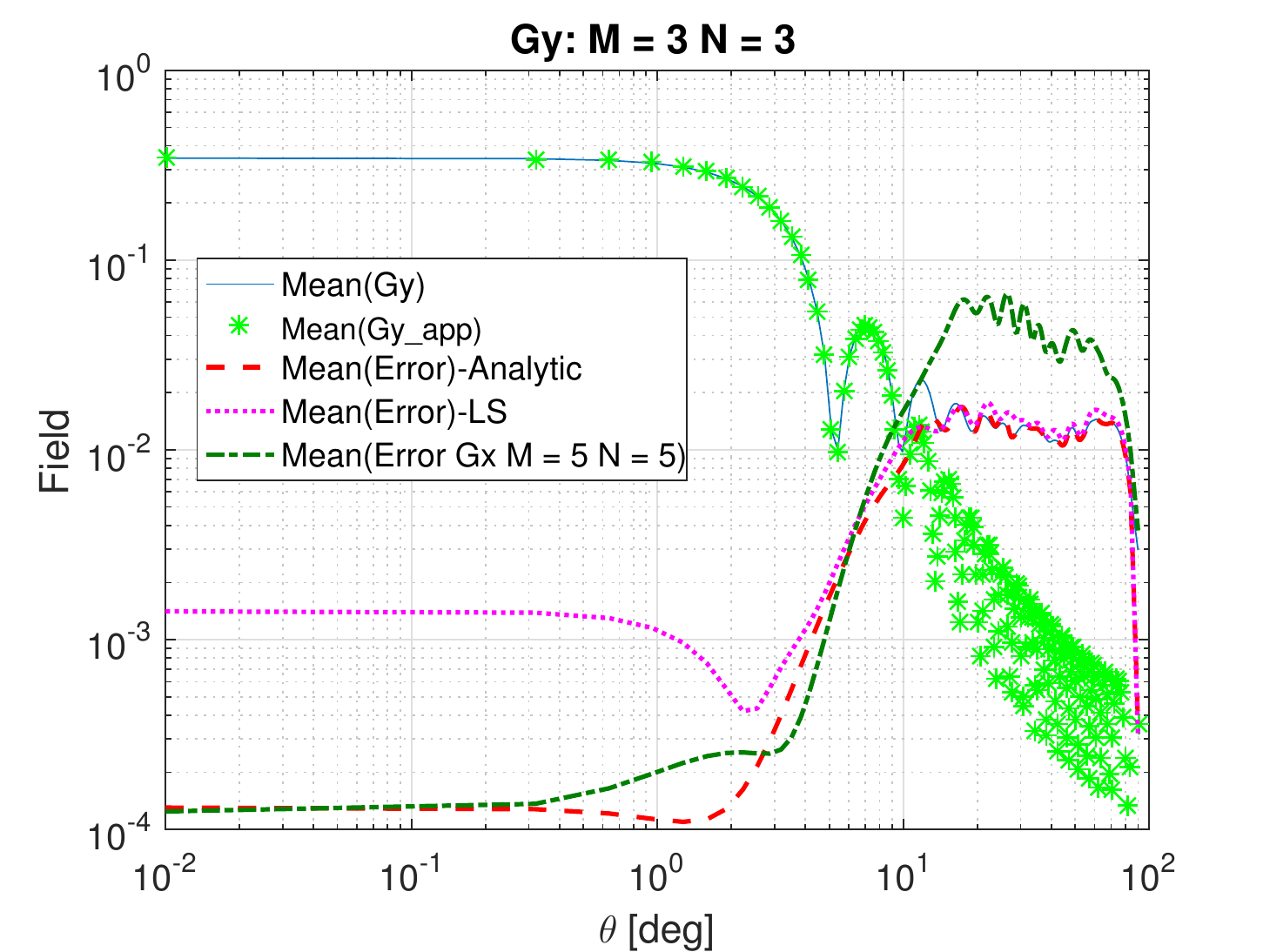}
	\caption{Mean of ${G_y}-$pattern over $\phi$ of the exact pattern (blue line), the modeled pattern (green-star) for (M=3, N=3), the error of the modeled ${G_y}-$pattern (dashed-pink/red line), and the mean of modeled ${G_x}-$pattern (dashed-dotted line) for (M=5,N=5).} \label{fig:py_cArray}
\end{figure}

For the cross-polarization component, ${G_y}$, fewer terms are required to achieve a comparable level of error w.r.t. ${G_x}$, as expected from the discussion in Section~\ref{sec:AA}. Fig.~\ref{fig:py_cArray} demonstrates this point, where the modeled beam of ${G_y}$ with $M=3$ and $N=3$ provides an error level as the modeled beam of ${G_x}$ with $M=5$ and $N=5$. As in this excitation configuration, ${G_x}$ is the co-polar component and ${G_y}$ is the cross-polar one, this fact will effectively reduce the number of coefficients required to model the full pattern.

\begin{figure}[!htb]
\centering
	\includegraphics[scale=0.65,clip,trim={0cm 0cm 0cm 0cm}]{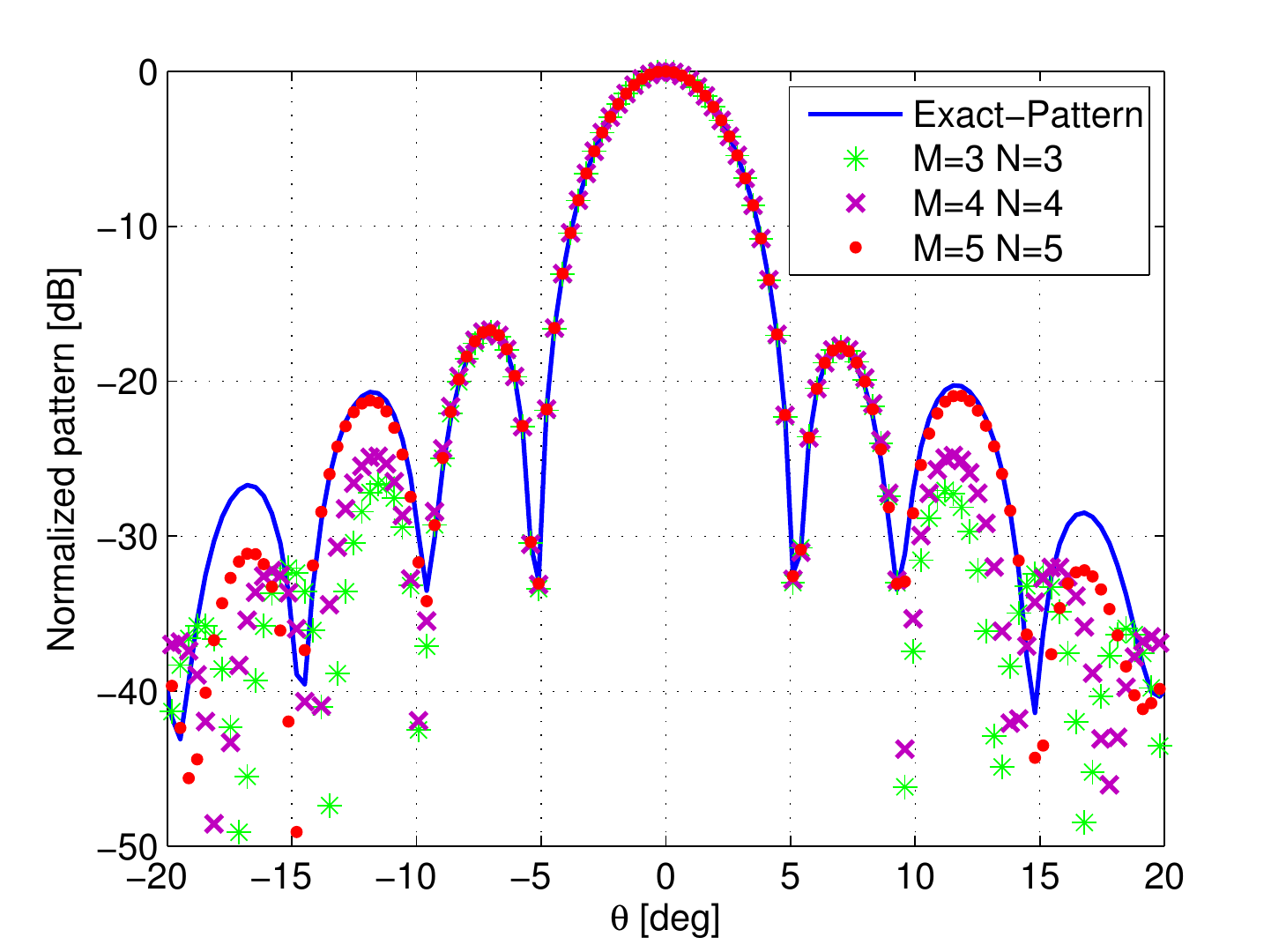}
	\caption{Radiation pattern of a SKA1-LOW station scanned at the broadside  ($\phi = 0^\circ$ cut) of the exact pattern (blue-line) using HARP, and modeled pattern for different values of (M,N) using the proposed approach.} \label{fig:rad_cArray}
\end{figure}

Finally, the modeled beam for the SKA1-LOW station is shown in Fig.~\ref{fig:rad_cArray} for different Zernike polynomial orders. The main beam and the first sidelobe are accurately modeled using only 56 coefficients (i.e. $M=N=3$ for each component). This validates the proposed method for the modeling of the SKA-LOW station.

\subsection{Scanned Array} \label{sec:SA}

\begin{figure}[!htb]
\centering
	\includegraphics[scale=0.6,clip,trim={0cm 0cm 0cm 0cm}]{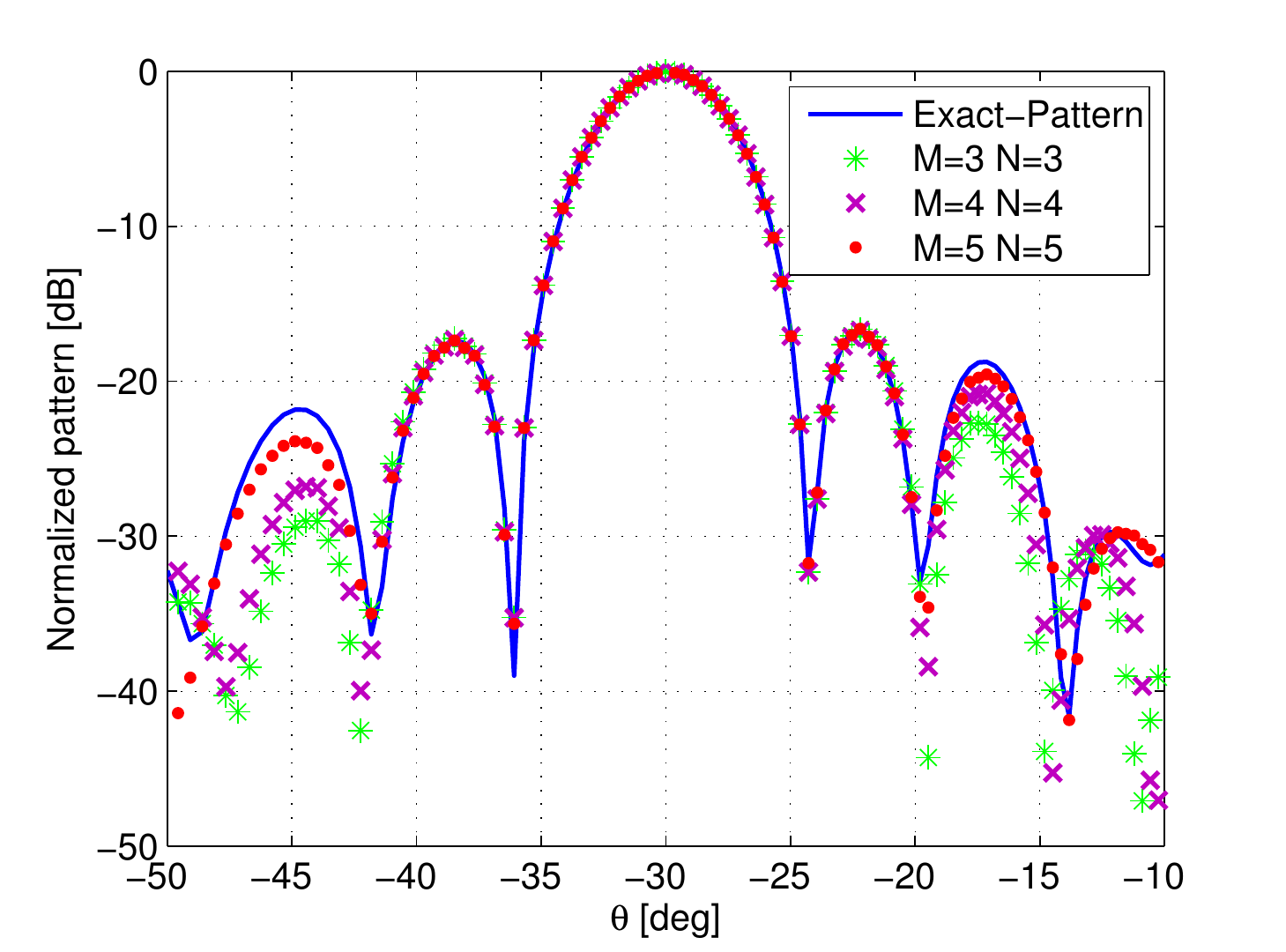}
	\caption{Radiation pattern of a SKA1-LOW station scanned at $(\phi,\theta) = (0,-30^\circ)$  of the exact pattern (blue-line) using HARP, and modeled pattern for different values of (M,N) using the proposed approach. ($\phi = 0^\circ$ cut).} \label{fig:rad_cArray_scan30}
\end{figure}

As presented in Section~\ref{subsec:arrayconfig}, we can exploit the shifting property of the Fourier transforms for the scanned array to effectively model the beam. Fig.~\ref{fig:rad_cArray_scan30} shows an example of the SKA1-LOW station scanned at $(\phi,\theta) = (0,-30^\circ)$, where the main beam of array pattern is successfully modeled using the proposed method. A similar performance is observed as for the case of the SKA1-LOW array scanned at broadside in Fig.~\ref{fig:rad_cArray}, i.e. a limited number of coefficients is needed to effectively represent the main beam and the first sidelobe.

\subsection{Different Array Shapes} \label{sec:DAS}

\begin{figure}[!htb]
\centering
\subfigure[Ellipse array configuration.]{\includegraphics[scale=0.375,clip,trim={0.25cm 0cm 1cm 0cm}]{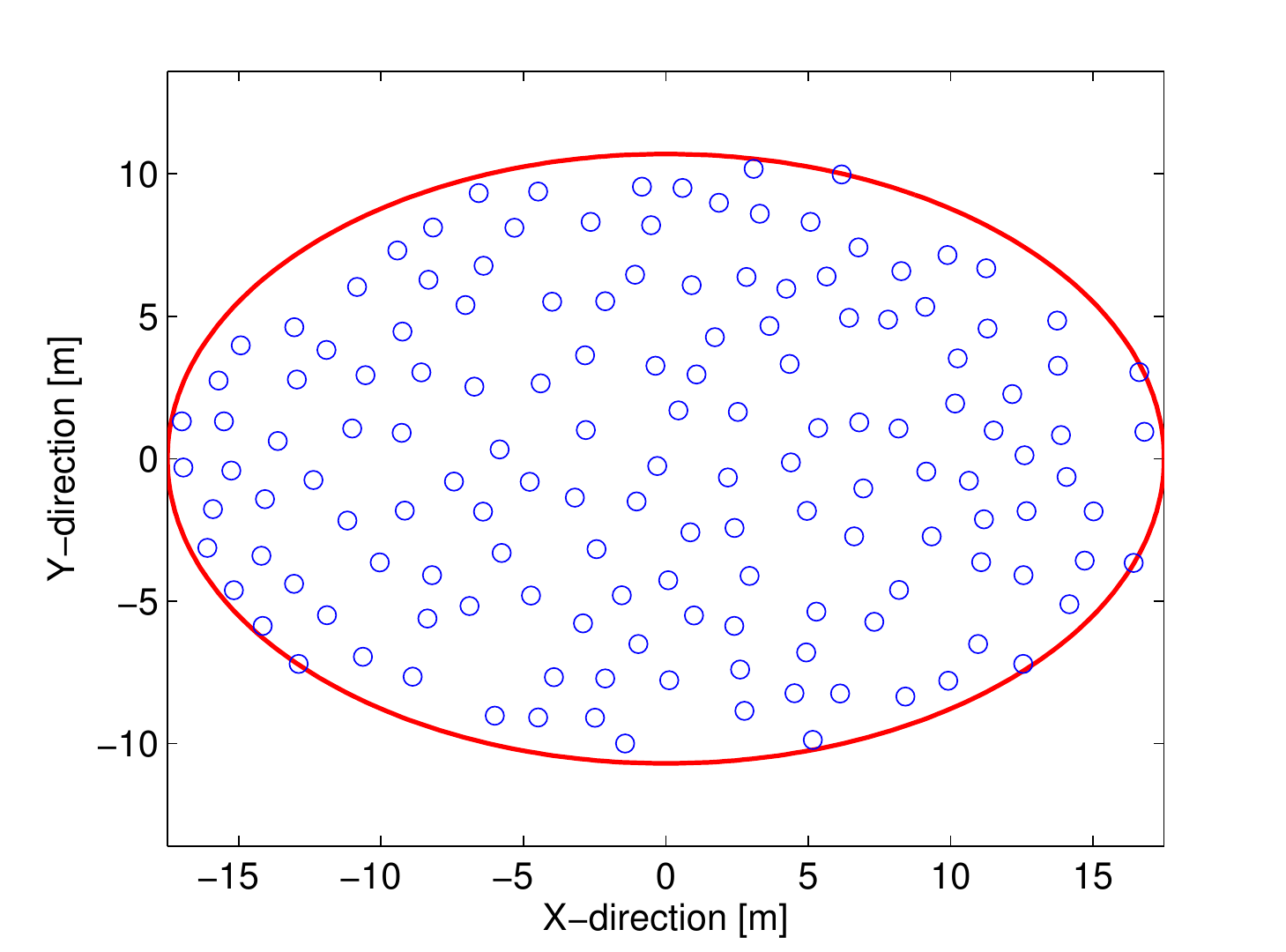} \label{fig:ellipse_conf}} 
\hspace*{1mm}
\subfigure[Mean ${G_x}$ with (M=3,N=3).]{\includegraphics[scale=0.375,clip,trim={0.25cm 0cm 1cm 0cm}]{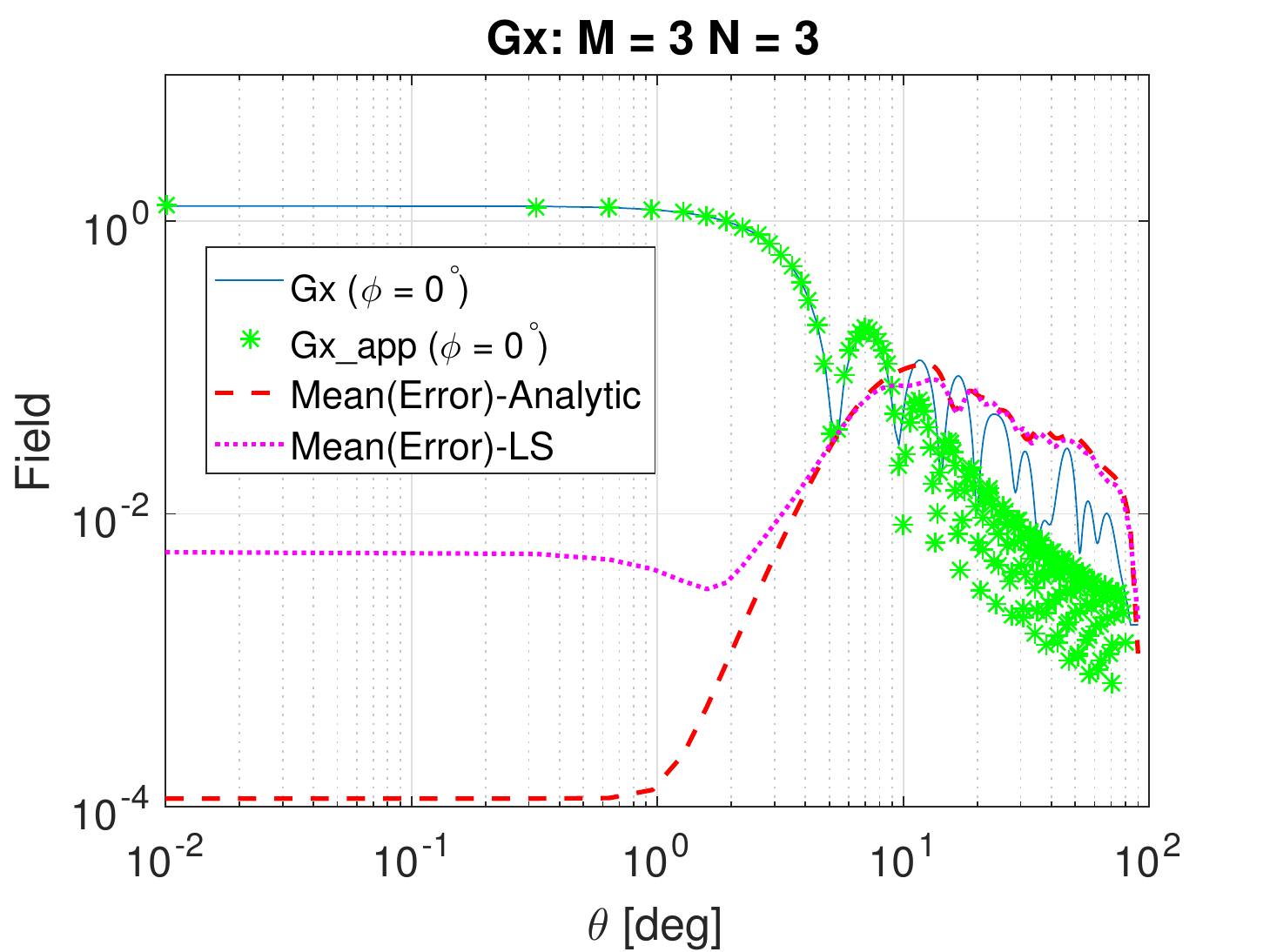} \label{fig:ellipse_3}} 
\hspace*{1mm}
	\subfigure[Mean ${G_x}$ with (M=4,N=4).]{\includegraphics[scale=0.375,clip,trim={0.25cm 0cm 1cm 0cm}]{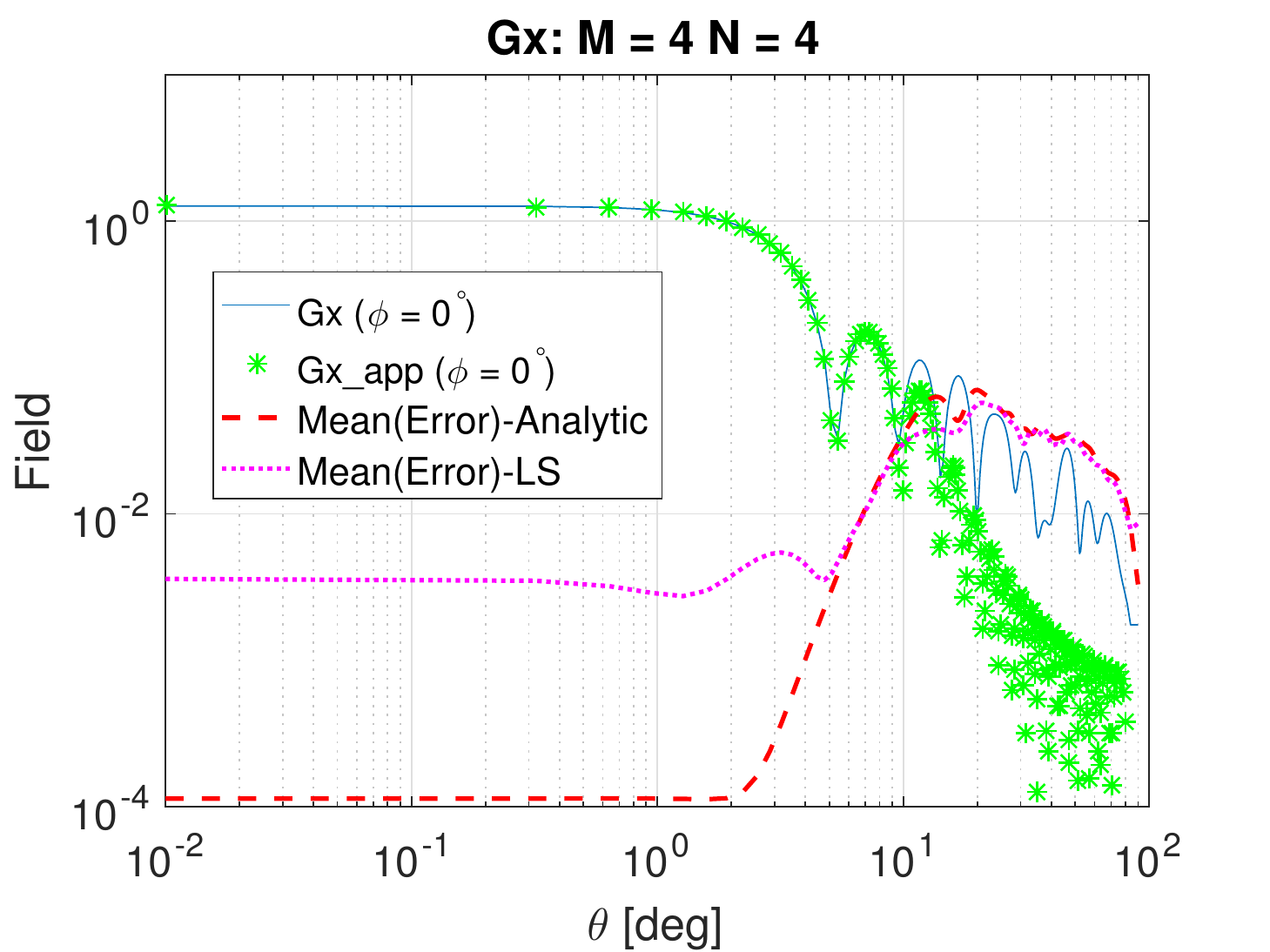}}
	\subfigure[Mean ${G_x}$ with (M=5,N=5).]{\includegraphics[scale=0.375,clip,trim={0.25cm 0cm 1cm 0cm}]{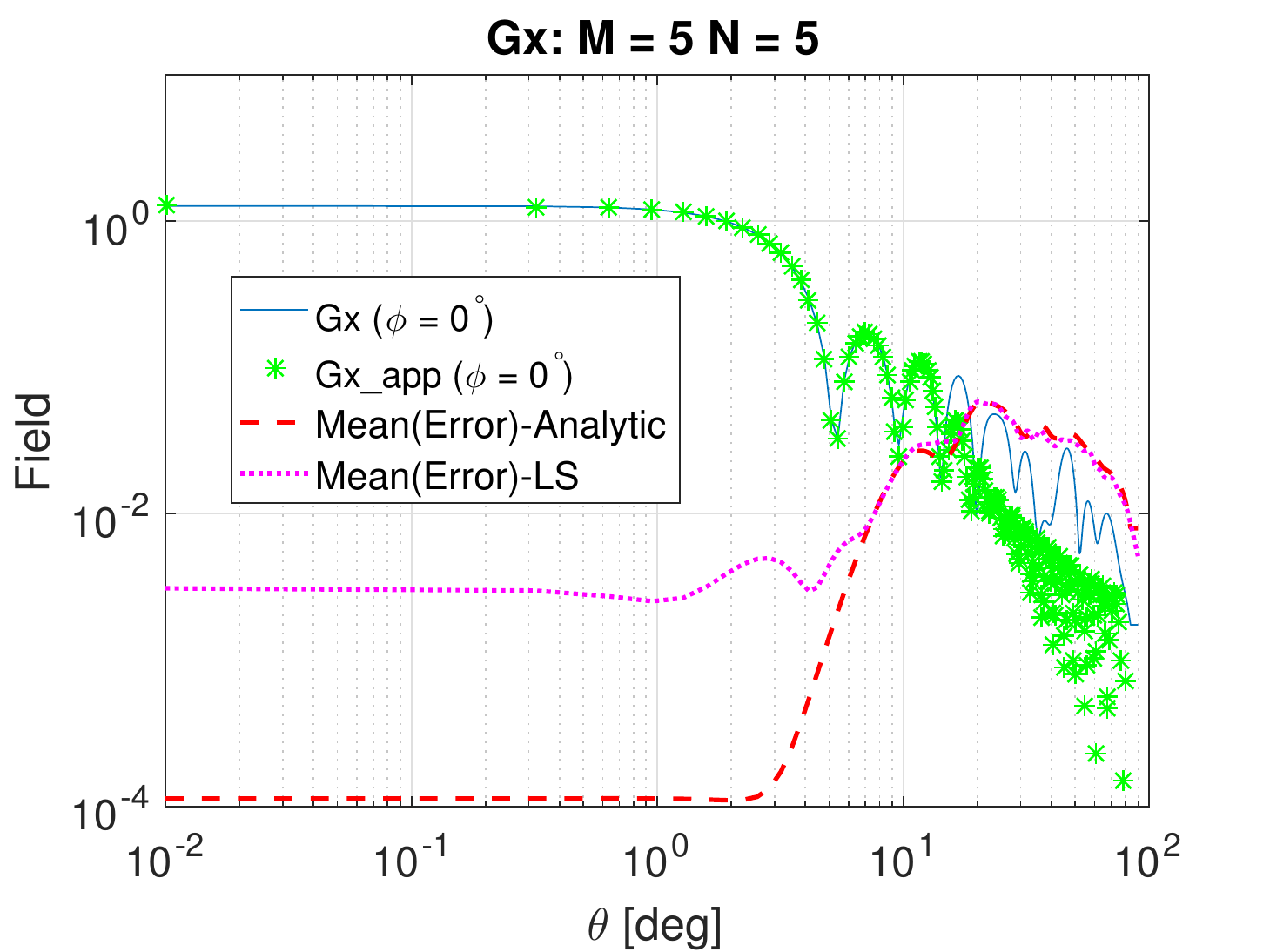}}
    
    \caption{Performance of the proposed approach to represent the array pattern of an ellipse array for different values of (M,N). (a) array of 146 SKALA over an ellipse surface; (b, c, d) $\phi = 0^\circ$ cuts for ${G_x}-$pattern of the exact and modeled patterns, and mean over $\phi$ for the difference between the exact and modeled patterns for (M=3,N=3), (M=4,N=4), (M=5,N=5), respectively. } \label{fig:ellipse}

\end{figure}

\begin{figure}[!htb]
\centering
\subfigure[Hexagon Array configuration.]{\includegraphics[scale=0.375,clip,trim={0.25cm 0cm 1cm 0cm}]{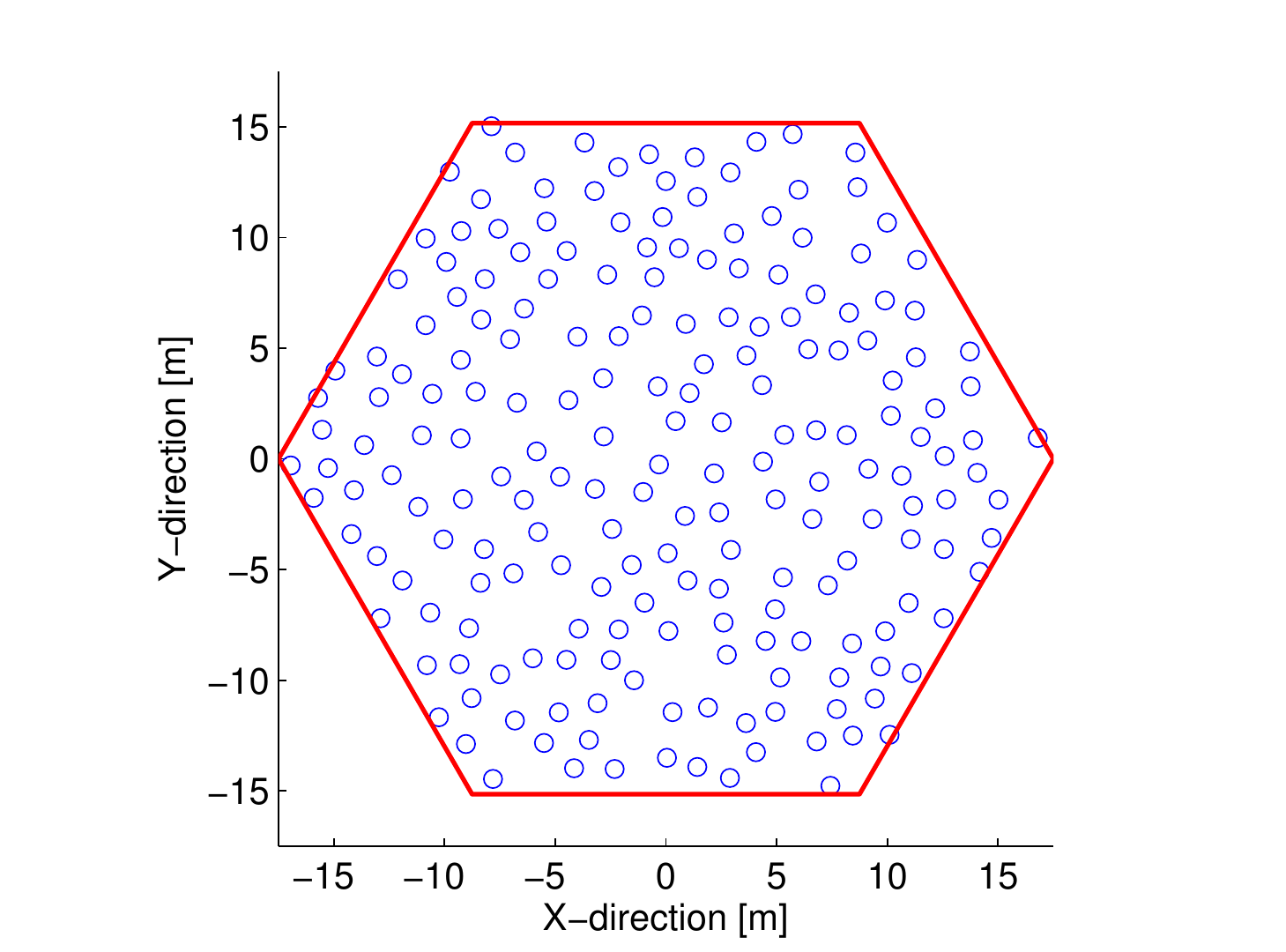} \label{fig:hexagon_conf}}  
\hspace*{1mm}
\subfigure[Mean ${G_x}$ with (M=3,N=3).]{\includegraphics[scale=0.375,clip,trim={0.25cm 0cm 1cm 0cm}]{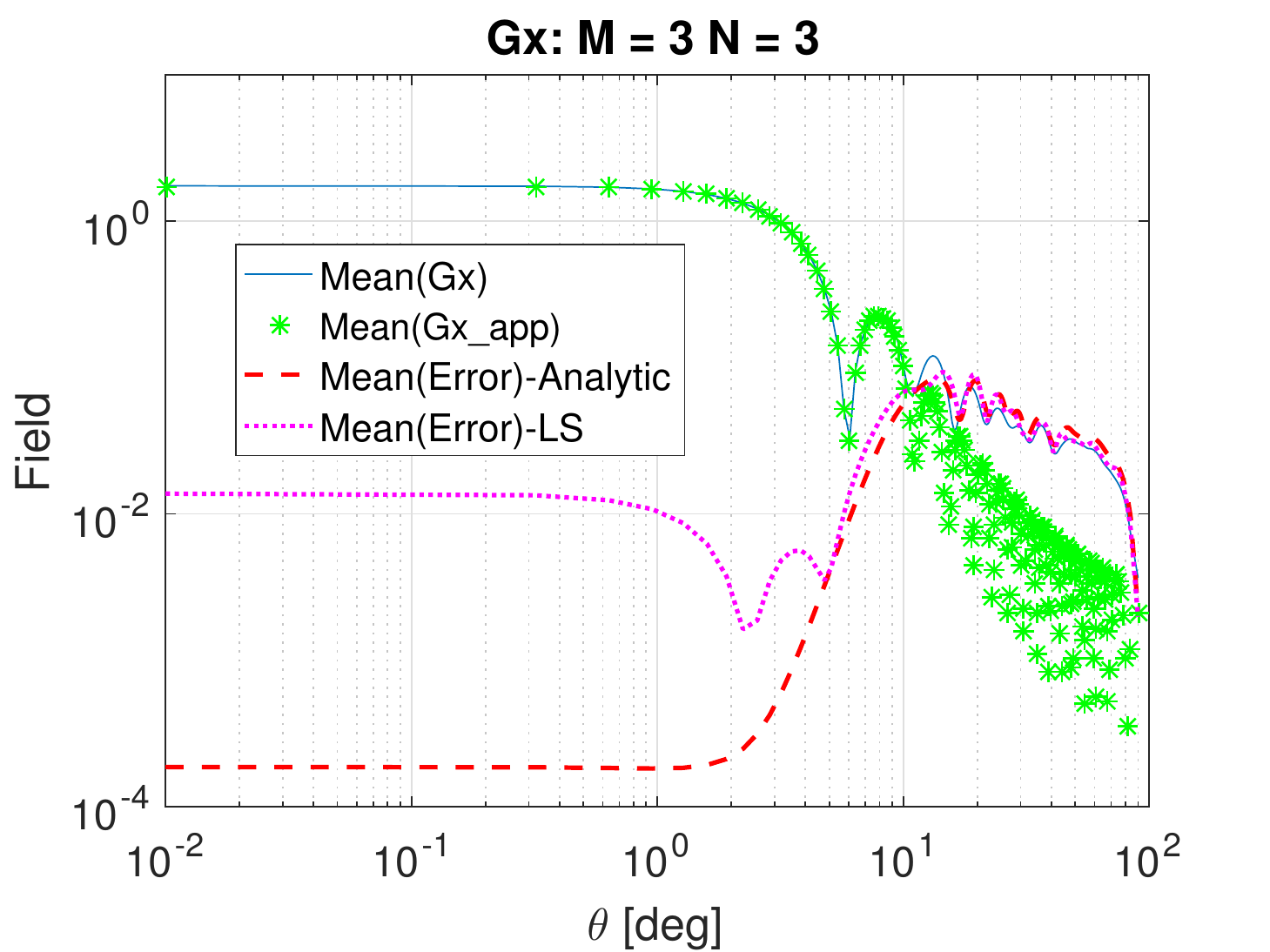} \label{fig:hexagon_3}}
\hspace*{1mm}
\subfigure[Mean ${G_x}$ with (M=4,N=4).]{\includegraphics[scale=0.375,clip,trim={0.25cm 0cm 1cm 0cm}]{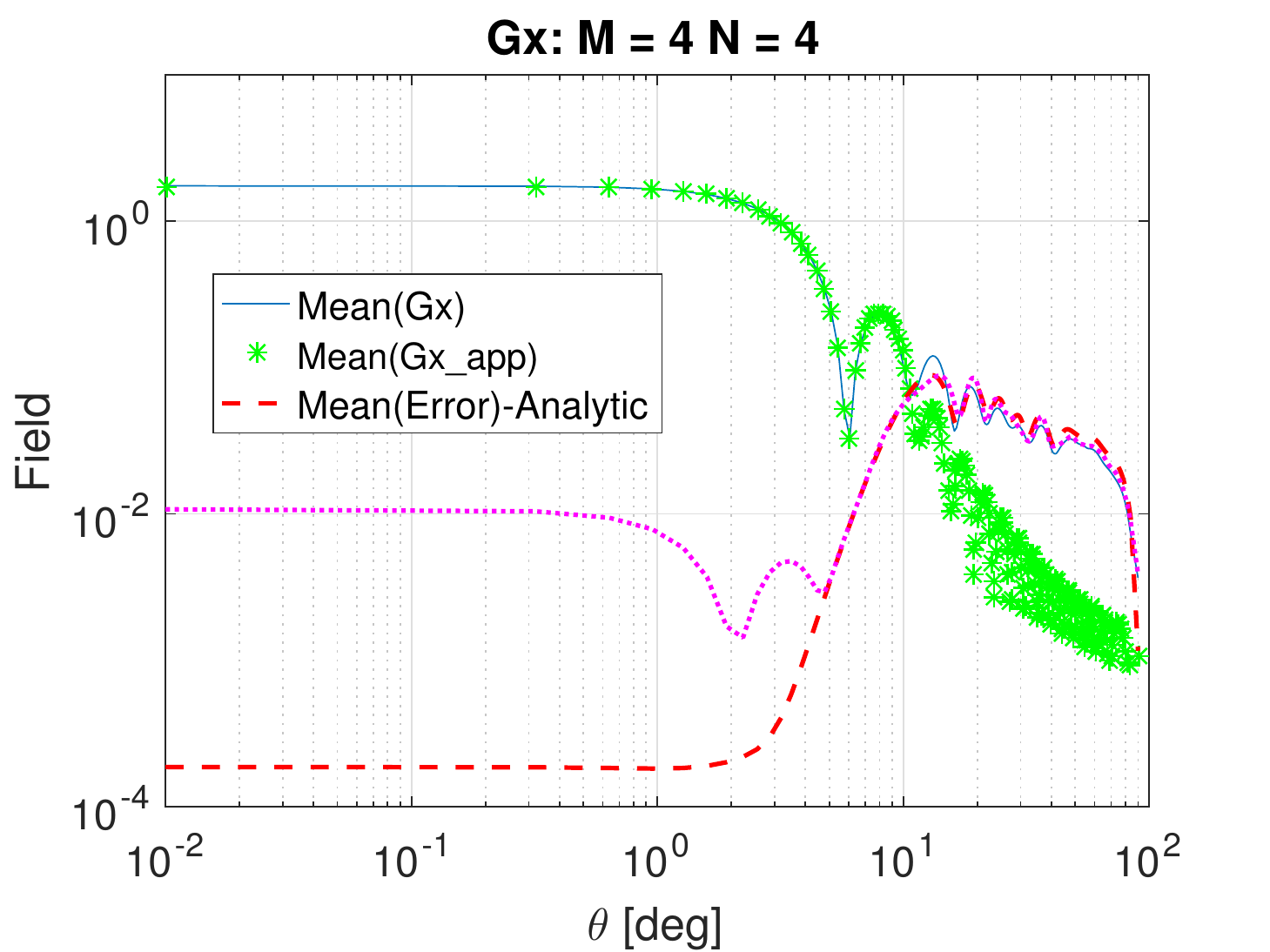}}
\subfigure[Mean ${G_x}$ with (M=5,N=5).]{\includegraphics[scale=0.375,clip,trim={0.25cm 0cm 1cm 0cm}]{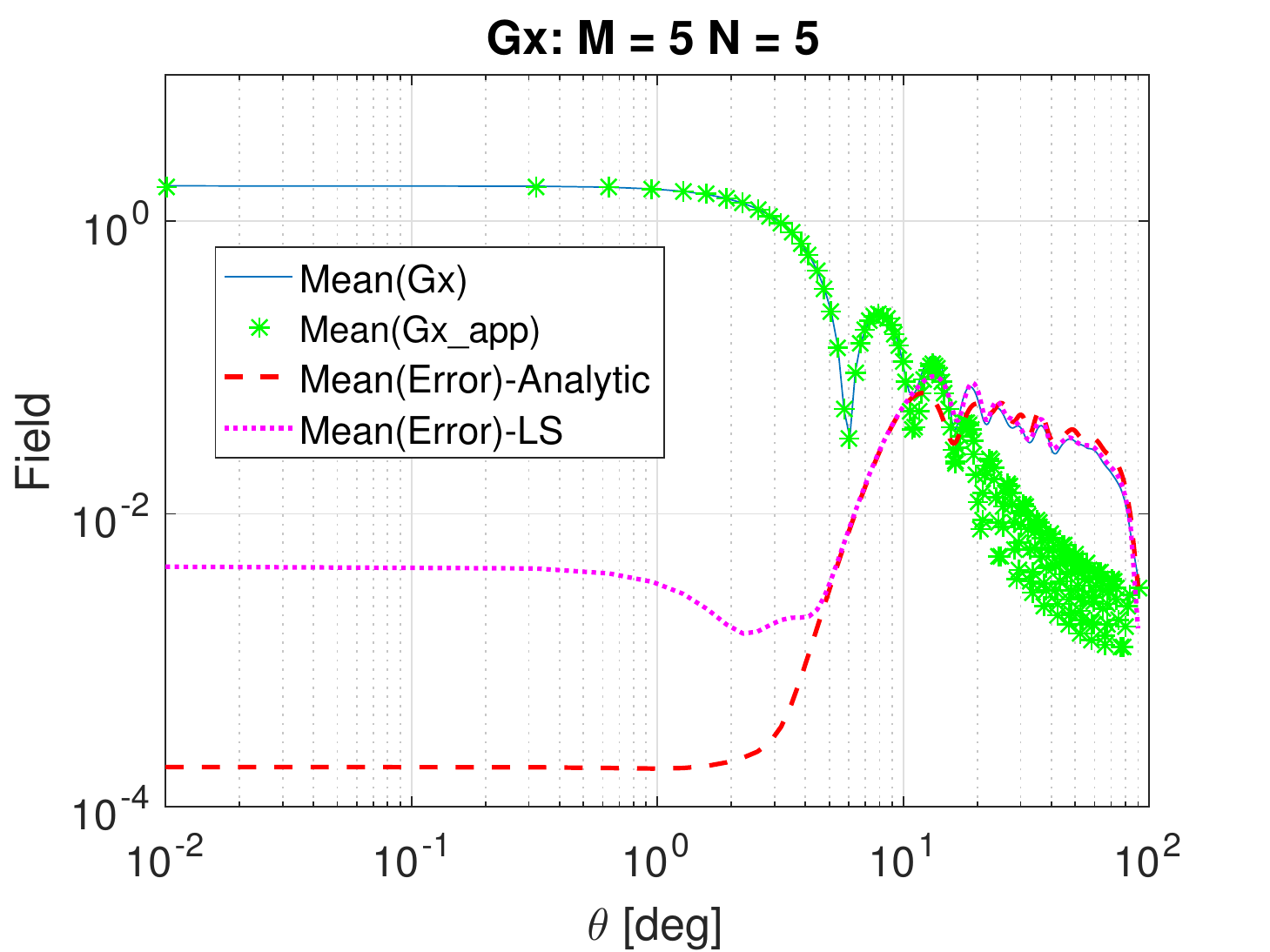}}
    
    \caption{Performance of the proposed approach to represent the array pattern of a hexagon array for different values of (M,N). (a) array of 201 SKALAs over a hexagon surface; (b, c, d) mean of ${G_x}-$pattern over $\phi$ of the exact, modeled patterns and their differences for (M=3,N=3), (M=4,N=4), (M=5,N=5), respectively. } \label{fig:hexagon}

\end{figure}

\begin{figure}[!htb]
\centering
\subfigure[Rectangular array configuration]{\includegraphics[scale=0.375,clip,trim={0.25cm 0cm 1cm 0cm}]{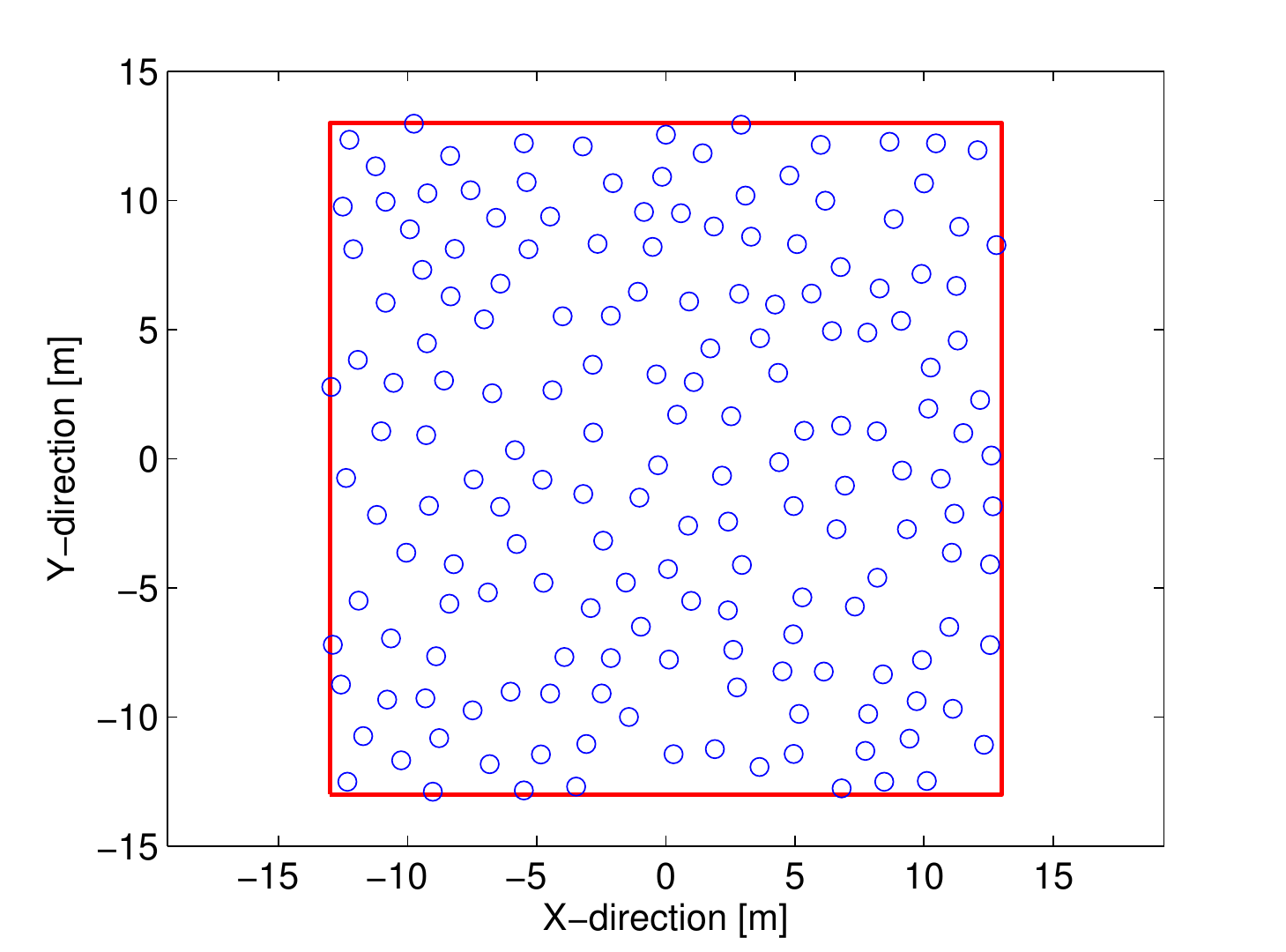}  \label{fig:rect_conf}} 
\hspace*{1mm}
\subfigure[Mean ${G_x}$ with (M=3,N=3).]{\includegraphics[scale=0.375,clip,trim={0.25cm 0cm 1cm 0cm}]{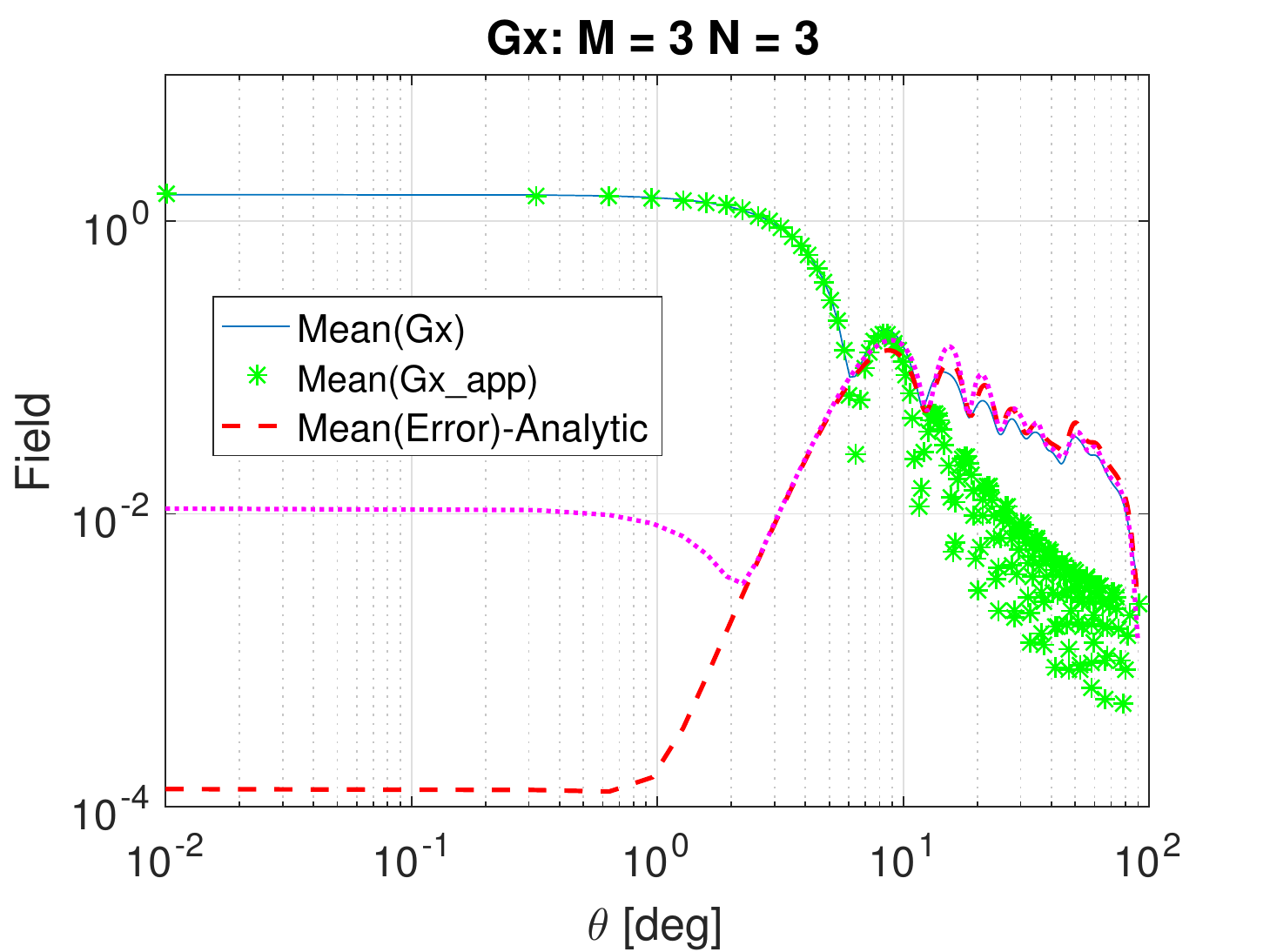} \label{fig:rect_3}} 
\hspace*{1mm}
\subfigure[Mean ${G_x}$ with (M=4,N=4).]{\includegraphics[scale=0.375,clip,trim={0.25cm 0cm 1cm 0cm}]{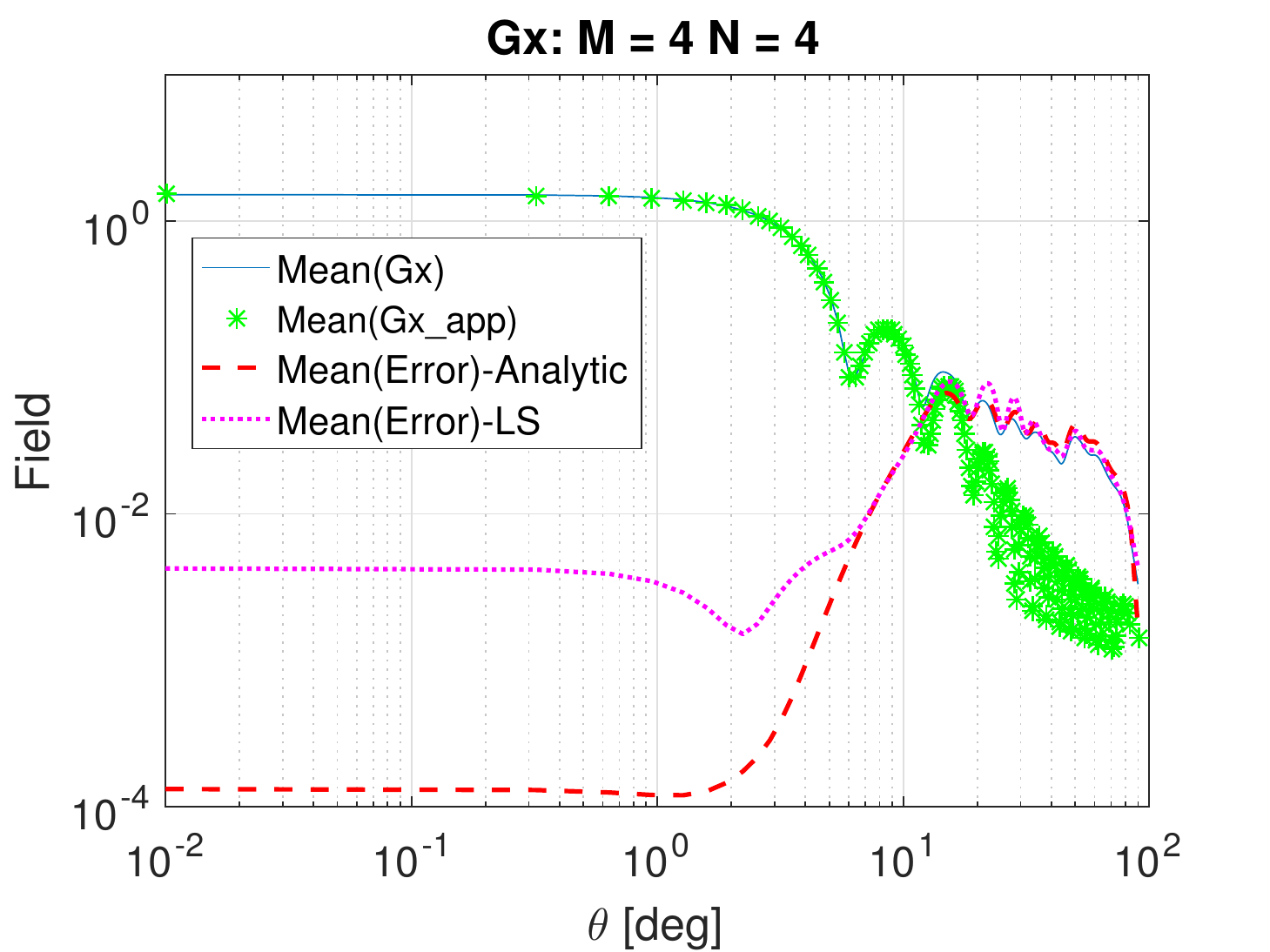}}
\subfigure[Mean ${G_x}$ with (M=5,N=5).]{\includegraphics[scale=0.375,clip,trim={0.25cm 0cm 1cm 0cm}]{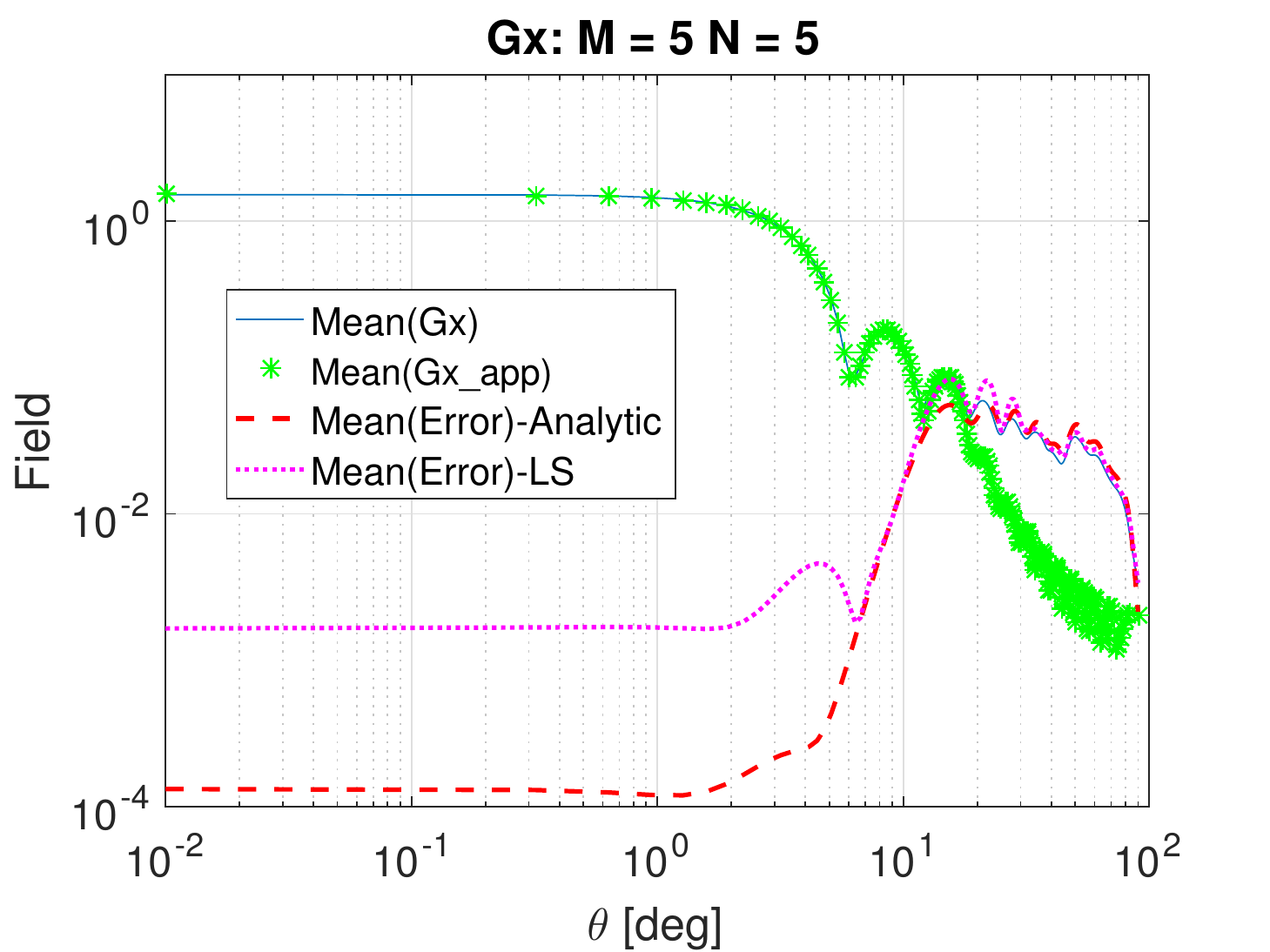}}
    
		\caption{Performance of the proposed approach to represent the array pattern of a rectangular array for different values of (M,N). (a) array of 175 SKALAs over a rectangular surface; (b, c, d) mean of ${G_x}-$pattern over $\phi$ of the exact, modeled patterns and their differences for (M=3,N=3), (M=4,N=4), (M=5,N=5), respectively.} \label{fig:rect}

\end{figure}

\begin{figure}[!htb]
\centering
\subfigure[Pentagon array configuration.]{\includegraphics[scale=0.375,clip,trim={0.25cm 0cm 1cm 0cm}]{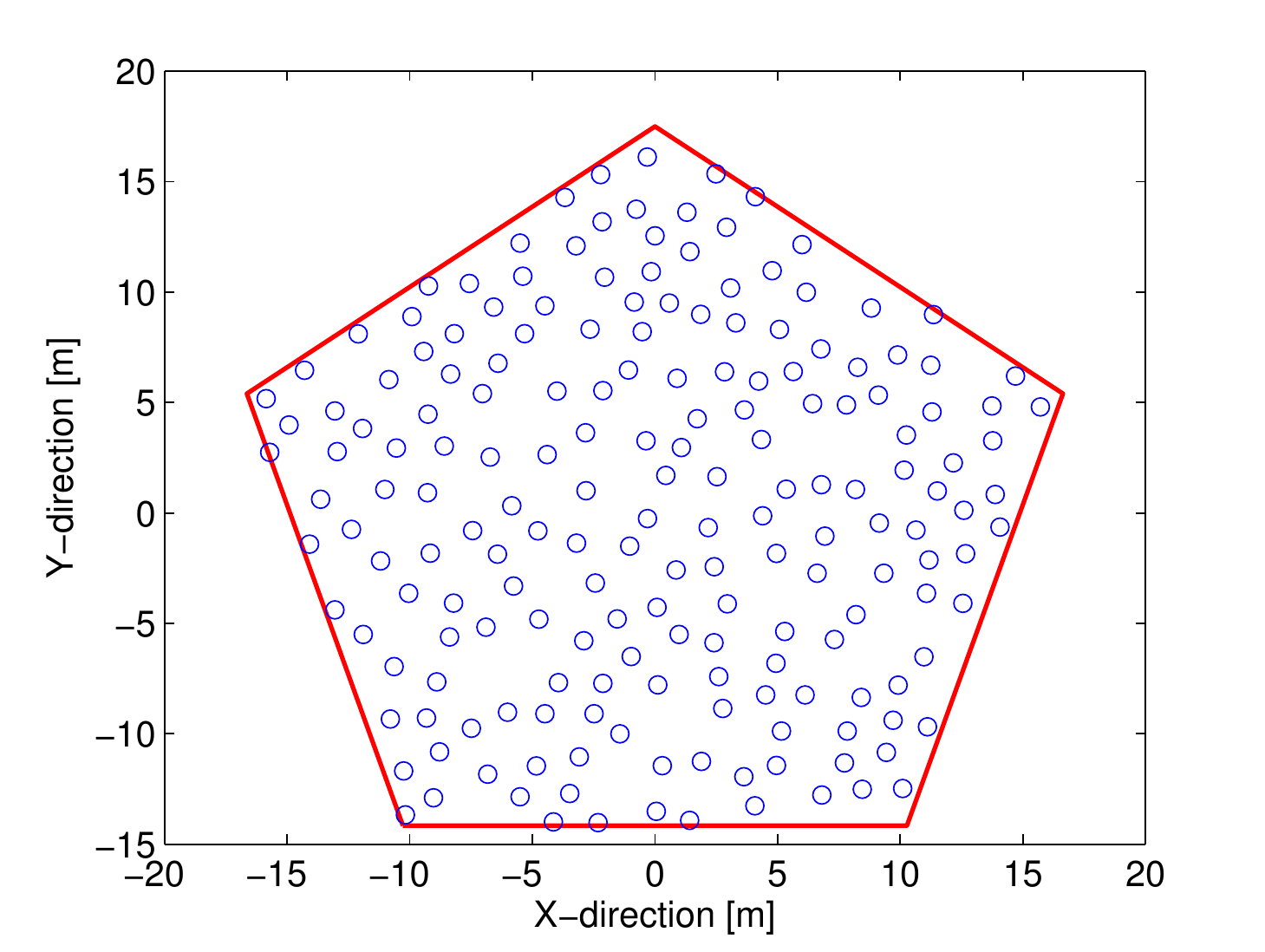} \label{fig:pentagon_conf}}
\hspace*{1mm}
\subfigure[Mean ${G_x}$ with (M=3,N=3).]{\includegraphics[scale=0.375,clip,trim={0.25cm 0cm 1cm 0cm}]{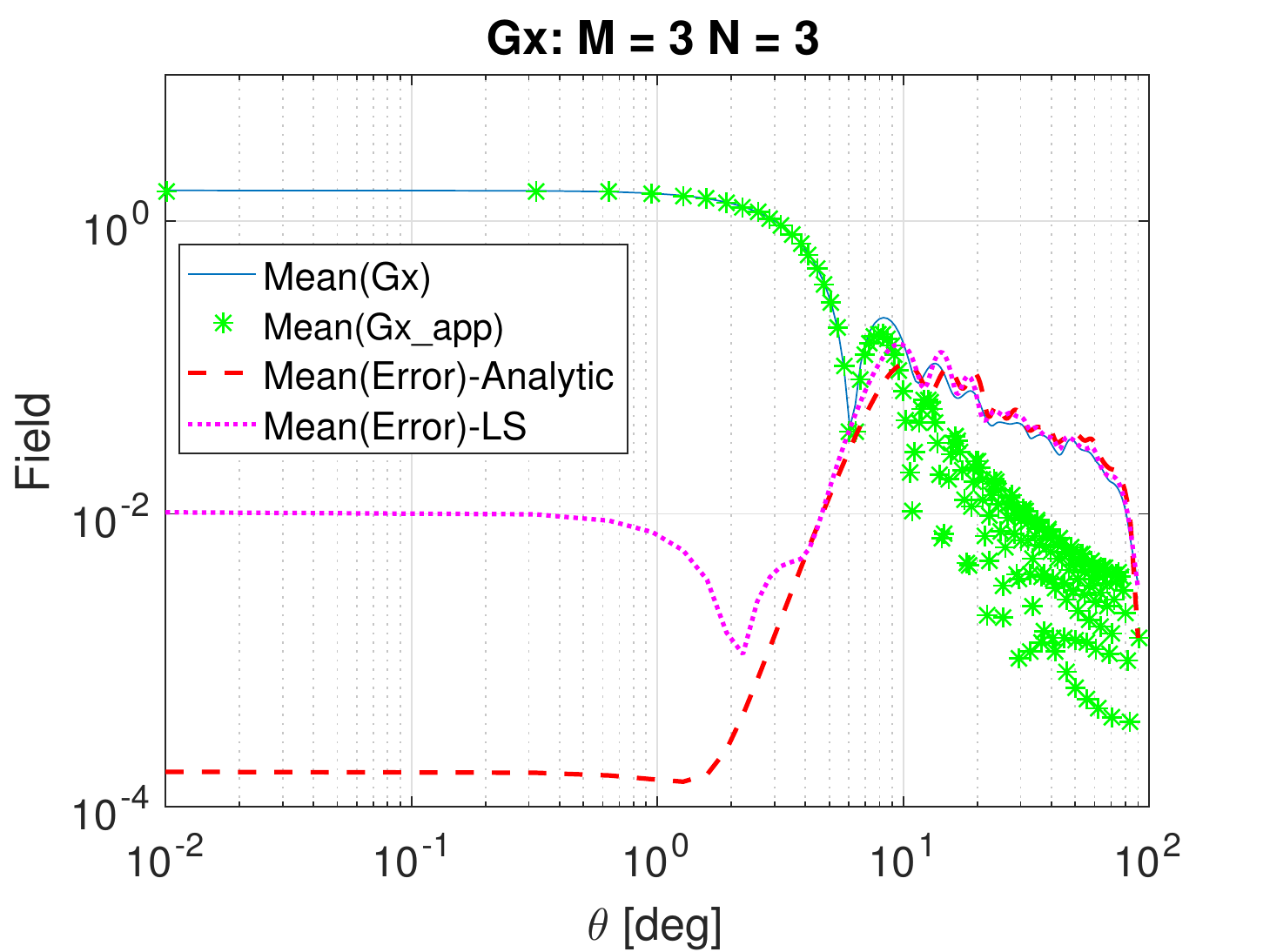} \label{fig:pentagon_3}} 
\hspace*{1mm}
\subfigure[Mean ${G_x}$ with (M=4,N=4).]{\includegraphics[scale=0.375,clip,trim={0.25cm 0cm 1cm 0cm}]{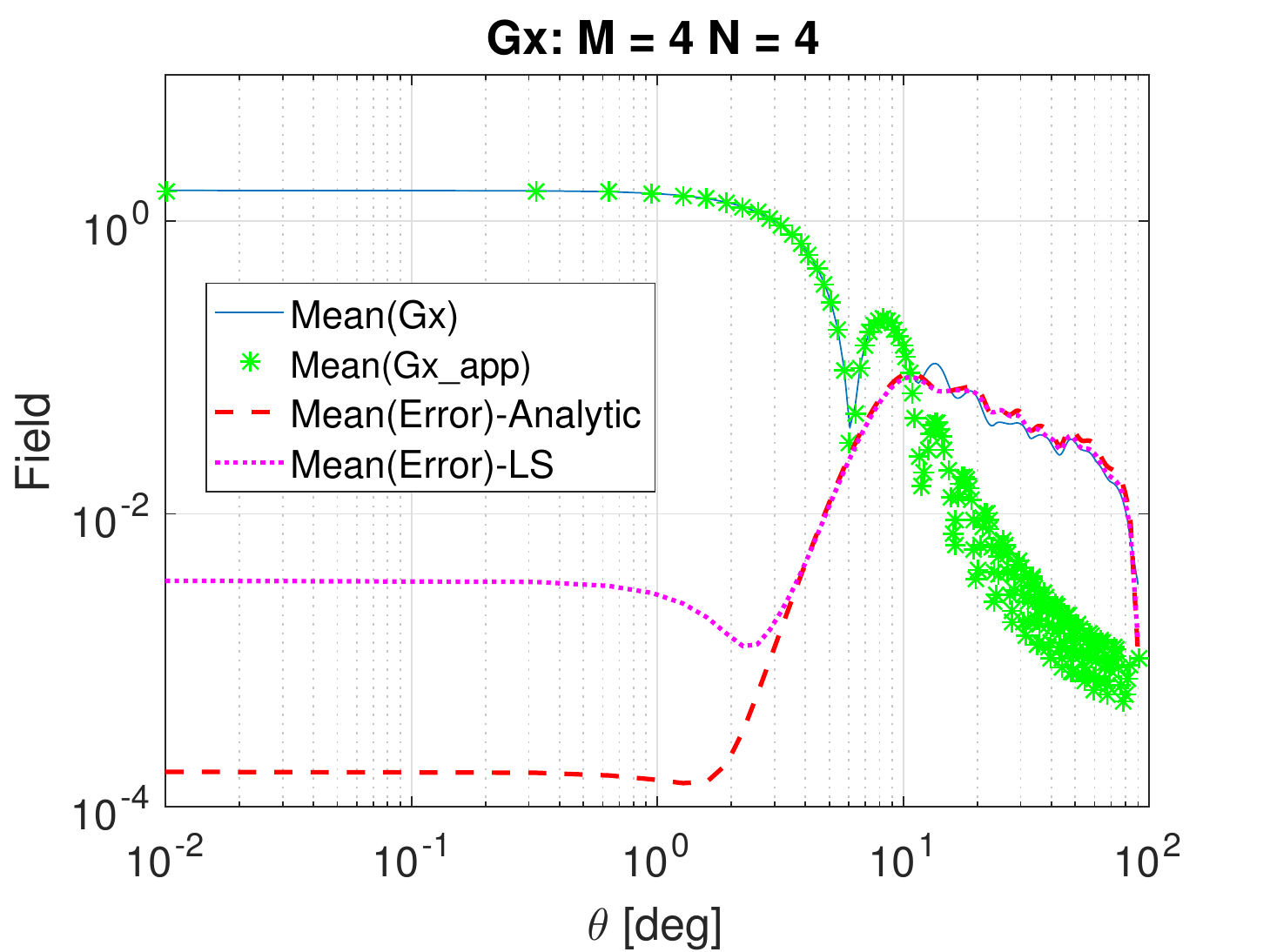}} 
\subfigure[Mean ${G_x}$ with (M=5,N=5).]{\includegraphics[scale=0.375,clip,trim={0.25cm 0cm 1cm 0cm}]{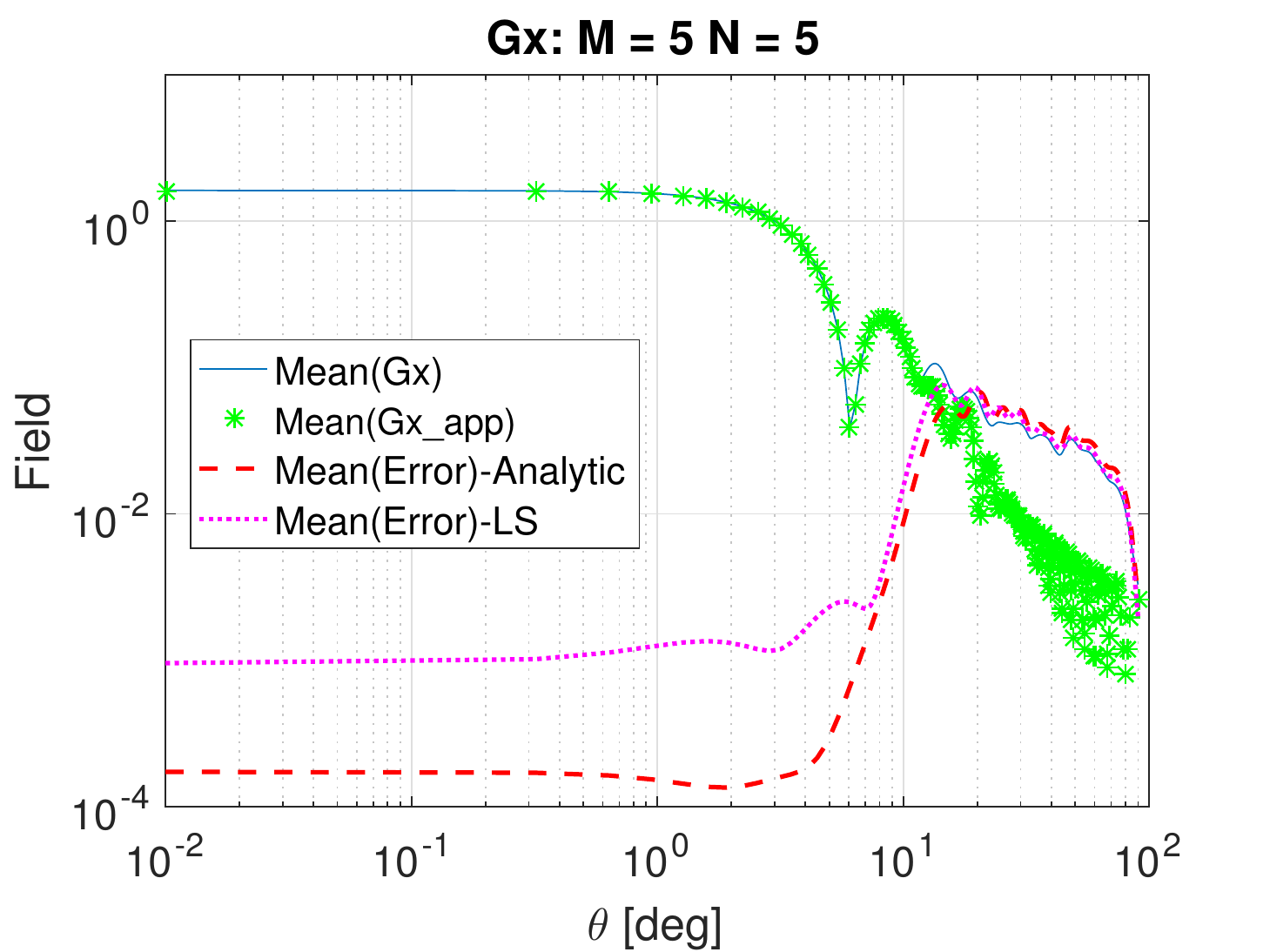}}
    
    \caption{Performance of the proposed approach to represent the array pattern of a pentagon array for different values of (M,N). (a) array of 186 SKALAs over a pentagon surface; (b, c, d) mean of ${G_x}-$pattern over $\phi$ of the exact, modeled patterns and their differences for (M=3,N=3), (M=4,N=4), (M=5,N=5), respectively.} \label{fig:pentagon}

\end{figure}

Different array configurations are considered in this section, as shown in Fig.~\ref{fig:ellipse}--Fig.~\ref{fig:pentagon}, including the ellipse, hexagon, rectangular and pentagon shapes with 146, 201, 175, and 186 SKALA elements, respectively. These array are constructed by selecting the elements of the circular SKA1-LOW station which are inside these shapes (or polygons). The patterns of these arrays are modeled using the proposed method, and the results are shown in Fig.\ref{fig:ellipse}--Fig.~\ref{fig:pentagon}.

Depending on the array shape, the modeled beam might require different numbers of coefficients to achieve a same level of accuracy. As shown in Fig.~\ref{fig:ellipse_3}~--~\ref{fig:pentagon_3}, the method barely modeled the main beam and first sidelobes for the rectangular, hexagon and pentagon with $M=3$ and $N=3$ (i.e. 28 coefficients). When the Zernike polynomials order increases to $M=4$ and $N=4$ (i.e. 45 coefficients), the first sidelobes are accurately modeled. The results indicate that for non-circular shape arrays, slightly higher orders might be needed for Zernike polynomials, but the method remains applicable.

One interesting point appears for the case of the elliptical array, for which the method exhibits a similar performance as for the circular array, thanks to the scaling property of the Fourier transforms. 
This result gives us another hint to deal with arrays, where the dimension in a given direction is very small compared to the orthogonal one, for example the flat ellipse or flat hexagon. 
Finally, it is worth commenting that for the other components of the pattern,  i.e. the cross-polarization, much fewer coefficients are required to achieve a same quality w.r.t the co-polarization level as discussed above.

\section{Conclusion}\label{sec:Concl}
An efficient approach to model the main beam and first few sidelobes of large irregular arrays used in radio astronomy (e.g. the SKA1-LOW telescope) has been presented. The array pattern is modeled as a function of the Fourier transforms of Zernike polynomials of different orders, where the coefficients are effectively determined by either an analytical approach or in the least-squares sense. Both provide similar performance at sidelobe, while the former one showed better performance in the main beam region. The results show that the main beam and first sidelobe can be accurately modeled by as few as 56 coefficients. The number of coefficients might vary for different array shapes. In general, the method is applicable for various array apertures such as the ellipse, hexagon, rectangular or pentagon shape arrays. The proposed method offers an effective means to model the array beam, when only the main beam and first sidelobe are considered.

\section*{Appendix I}
We consider a 2D aperture in the $xy-$plane with an equivalent current distribution $I^e \equiv (I^e_x,I^e_y)$, with a Fourier transform given by $f_t \equiv (f_{tx},f_{ty})$. The radiation pattern $G \equiv (G_x,G_y,G_z)$ of the aperture can be expressed as the following projection~\cite{Kildal15}
\begin{equation}
\bar{G} = C \Big( \bar{f_t} - \hat{u} (\bar{f_t}.\hat{u}) \Big)  \label{eq:p_f}
\end{equation}
where $C = {-jk\eta}\big/{4\pi}$ with $\eta$ is the free-space impedance. The inverse Fourier transform of $G_x$ corresponds to the ``pseudo-current'' $I_x$ (see Equation~(\ref{eq:current})). Likewise $I_y$ and $G_y$ are linked by a Fourier transform.  From~(\ref{eq:p_f}), the $G_x$ and $G_y$ components of the pattern are obtained as
\begin{align}
G_x &= C \Big(f_{tx} - u_x(f_{tx}u_x + f_{ty}u_y) \Big)  \label{eq:px} \\
G_y &= C \Big(f_{ty} - u_y(f_{tx}u_x + f_{ty}u_y) \Big) \label{eq:py}
\end{align}
It is noted that the $G_z$ pattern is automatically obtained from $G_x$ and $G_y$ patterns via the relation $\bar{G}.\hat{u} = G_xu_x + G_yu_y + G_zu_z = 0$. The above expression is consistent with  Equation (11) in~\cite{Ludwig73}, noting that $\hat{u} \equiv (u_x,u_y,u_z) = (\sin\theta \cos\phi,\sin\theta \sin\phi,\cos\theta)$.

From~(\ref{eq:px}) and~(\ref{eq:py}), one can easily express $(f_{tx},f_{ty})$ in terms of $(G_x,G_y)$ as
\begin{equation}    
\begin{bmatrix} f_{tx} \\ f_{ty} \end{bmatrix}
=
\frac{1}{{C\,u^2_z}} \begin{bmatrix} 1-{u^2_y}     \,\,\,\,\,\,\,\,\, {u_xu_y}   \\ \,\, {u_xu_y} \,\,\,\,\,\,\,\,\, \,\,\,  1-{u^2_x}   \end{bmatrix} \begin{bmatrix}
    G_x \\
    G_y \\
\end{bmatrix}
\label{eq:f_p}
\end{equation}
The left-hand side of~(\ref{eq:f_p}) is the Fourier transform of equivalent currents, while $[G_x \,\,G_y]^T$ is the Fourier transform of the ``pseudo current'' used in this paper. Eq.~(\ref{eq:f_p}) provides the link between these two currents. It would also be possible to start from physical or equivalent currents (if available), and to model the pattern $(f_{tx}, f_{ty})$ using the technique presented in this paper. The radiation pattern $G_x,G_y$ are then obtained from (\ref{eq:px}) and~(\ref{eq:py}). For arrays of complex antennas with the available radiation pattern, the ``pseudo-current'' probably is the most straightforward way, because it only involves a Fourier transform link without any projection.

As a reminder, equivalent electric $\bar{I}^e$ and magnetic currents $\bar{M}^e$~\cite[Section 4.3]{Kildal15},~\cite{Harrington01}, on the surface enclosing electromagnetic sources correspond to $\hat{n}\times\bar{H}$ and $\bar{E}\times\hat{n}$, where $\bar{E}$ and $\bar{H}$ are electric and magnetic fields, $\hat{n}$ is the normal unit vector to the surface. When the surface corresponds an infinite plane, the magnetic equivalent current $\bar{M}^e$ can be omitted if the equivalent electric currents are doubled~\cite{Harrington01}.

\section*{Acknowledgment}

The authors would like to thank Quentin Queuning for the mesh of SKALA, David Gonz\'alez-Ovejero for initiating the interpolatory method, and Michel Arts at ASTRON, the Netherlands for providing the WIPL-D simulation data. This research was supported by the Science \& Technology Facilities Council (UK) grant: \textit{SKA, ST/M001393/1} and the University of Cambridge, UK.


\end{document}